\documentclass[final,3p,times]{elsarticle}

\usepackage{amssymb}
\usepackage{amsmath}
\usepackage{float}
\usepackage[hidelinks]{hyperref}  % or \usepackage[colorlinks]{hyperref}
\usepackage{mathrsfs}
\usepackage[para,online,flushleft]{threeparttable}
\usepackage{multirow}% for table
\usepackage{makecell}% for table
\usepackage{colortbl}% for table
\usepackage{xcolor}% for table
\usepackage{array}% for table
\usepackage{adjustbox}% for table
\usepackage{booktabs}% for table
\usepackage{tabularx}  % for auto-width columns

\journal{Applied Mathematical Modelling}

\begin{document}

\begin{frontmatter}

\title{Reconstruction of three-dimensional shapes of normal and disease-related erythrocytes from partial observations using multi-fidelity neural networks}

\author[a]{Haizhou Wen}
\author[b]{He Li}
\author[a]{Zhen Li\corref{cor1}}

\cortext[cor1]{Corresponding Author. Email: zli7@clemson.edu}

\affiliation[a]{organization={Department of Mechanical Engineering},
            addressline={Clemson University},
            city={Clemson},
            state={SC 29634},
            country={USA}}
\affiliation[b]{organization={School of Chemical, Materials, and Biomedical Engineering},
            addressline={University of Georgia},
            city={Athens},
            state={GA 30602},
            country={USA}}

%% Abstract
\begin{abstract}
Reconstruction of three-dimensional (3D) erythrocyte or red blood cell (RBC) morphology from partial observations, such as microscope images, is essential for understanding the physiology of RBC aging and the pathology of various RBC disorders. In this study, we propose a multi-fidelity neural network (MFNN) approach to fuse high-fidelity cross-sections of an RBC, with a morphologically similar low-fidelity reference 3D RBC shape to recover its full 3D surface. The MFNN predictor combines a convolutional neural network trained on low-fidelity reference RBC data with a feedforward neural network that captures nonlinear morphological correlations, and augments training with surface area and volume constraints for regularization in the low-fidelity branch. This approach is theoretically grounded by a topological homeomorphism between a sphere and 3D RBC surfaces, with training data generated by dissipative particle dynamics simulations of stomatocyte–discocyte–echinocyte transformation. Benchmarking across diverse RBC shapes observed in normal and aged populations, our results show that the MFNN predictor can reconstruct complex RBC morphologies with over 95\% coordinate accuracy when provided with at least two orthogonal cross-sections. It is observed that informative oblique cross-sections intersecting spicule tips of echinocytes improve both local and global feature reconstruction, highlighting the value of feature-aware sampling. Our study further evaluates the influence of sampling strategies, shape dissimilarity, and noise, showing enhanced robustness under physically constrained training. Altogether, these results demonstrate the capability of MFNN to reconstruct the 3D shape of normal and aged RBCs from partial cross-sections as observed in conventional microscope images, which could facilitate the quantitative analysis of RBC morphological parameters in normal and disease-related RBC samples.
\end{abstract}

%% Keywords
\begin{keyword}
Machine learning, Neural network, Cell aging, Red blood cell, Stomatocyte-Discocyte-Echinocyte  
\end{keyword}

\end{frontmatter}

\section{Introduction}
The morphology of the erythrocyte, or red blood cell (RBC), is a vital indicator of the functionality of RBCs in circulation. The highly deformable, naturally biconcave shape enables RBC to undergo large deformation under external forces, thereby ensuring the efficient transport of oxygen and essential nutrients throughout the bloodstream~\cite{annurev:/content/journals/10.1146/annurev.fluid.37.042604.133933}. However, RBC can undergo reversible or irreversible shape transformations under different physiological conditions~\cite{wong2018explanation}, influenced by various physical or chemical factors such as the concentration of positive or negative ionic lipids~\cite{sheetz1974biological0}, pH value~\cite{kuzman2000effect}, cholesterol levels~\cite{https://doi.org/10.1002/jss.400080404}, adenosine triphosphate (ATP) concentration~\cite{doi:10.1073/pnas.0904614106}, and contact with glass surfaces~\cite{ERIKSSON1990193}. The morphological transformation of RBC is often accompanied by changes in macroscopic mechanical properties like cell stiffness~\cite{xu2018stiffness} and blood viscosity~\cite{doi:10.1073/pnas.0904614106}, leading to a variety of normal and abnormal phenomena such as blood metabolism~\cite{balanant2018experimental} and blood disorders. Numerous and extensive studies employing experiments and numerical simulations have been investigating the relation between morphology transformations and blood disorders caused by pathological factors, such as malaria~\cite{doi:10.1073/pnas.1505584112}, sickle cell anemia (SCA)~\cite{10.1172/JCI106273}, hereditary spherocytosis or elliptocytosis~\cite{li2012two}, diabetes~\cite{BROWN2005295}, hemolysis~\cite{pan2022fatigue}, and thrombosis induced by COVID-19~\cite{10.7554/eLife.81316}. Decades of research have revealed an association between the morphology and mechanical properties of RBCs, highlighting the critical role of RBC geometry in various physiological conditions. Moreover, numerous numerical studies~\cite{pozrikidis2003numerical,annurev:/content/journals/10.1146/annurev-fluid-010313-141349,li2017computational,tang2017openrbc,10.1063/5.0050747,enjalbert2024effect} have demonstrated that accurately representing RBC geometry across different morphologies is a crucial prerequisite for defining the initial configuration in simulations, which in turn determines the reliability of the simulation results. Therefore, capturing an accurate and well-configured three-dimensional (3D) geometry of RBCs is a fundamental step toward uncovering the biophysical mechanisms underlying the morphology-property correlation.

The conventional approach to directly obtain cell geometry is to use optical imaging techniques, which translate biological cells and subcellular structures into visible images and have been extensively developed in physics laboratories over the decades. Among these, quantitative phase imaging (QPI)~\cite{s130404170} is a label-free, non-invasive optical microscopy technique that has been actively employed and further advanced in the field of cellular biology. By measuring the phase shifts in light transmitted through transparent specimens, QPI generates quantitative optical path length maps of the sample, thereby enabling the reconstruction of its 3D geometry~\cite{Cuche:99}. Building upon QPI, in the 1960s, Wolf~\cite{WOLF1969153} first introduced the concept of optical diffraction tomography (ODT), which extends QPI’s capabilities by acquiring multiple phase images from various illumination angles and reconstructing the sample’s 3D refractive index (RI) distribution through inverse scattering algorithms. However, ODT was not applied to 3D cellular imaging until 2009~\cite{Sung:09}. Since then, numerous implementations~\cite{Haeberle20052010,10.1371/journal.pone.0049502} of ODT have been developed and applied, including 3D imaging of RBCs~\cite{Kim:13}. Tomographic phase microscopy (TPM)~\cite{Jin:17} represents one such implementation. Early TPM systems typically relied on incident laser illumination angle scanning using galvanometer mirrors or sample rotation to obtain projection field measurements~\cite{charriere2006cell}. With advances in modern computational hardware, TPM performance has significantly improved in terms of temporal and spatial resolution~\cite{kus2015active} as well as graphics processing speed~\cite{Dardikman:16}, making it an increasingly popular tool in cellular biophysics research, including studies on the hematology of RBCs~\cite{10.1002/jbio.201600113} and white blood cells (WBCs)~\cite{Yoon:15}, blood disorders caused by parasites~\cite{10.1073/pnas.0806100105}, and the microstructure of subcellular organelles~\cite{kim2016three}. Overall, TPM and ODT are reliable and widely used techniques for high-resolution, label-free 3D imaging of live cells. However, as conventional tomographic methods, such as TPM and ODT, are increasingly applied to complex biological scenarios, a number of challenges emerge. On the one hand, accurate measurement relies on selecting a suitable algorithm to capture the RI distribution for a given scenario. This often requires developing or modifying a new algorithm tailored to each specific case, thereby limiting the generalizability of any single approach. On the other hand, more sophisticated algorithms designed to handle complex biological structures and achieve higher-resolution reconstructions place greater demands on hardware resources, resulting in increased computational cost. Therefore, how to extend the generalizability and reduce the computational burden of conventional approaches remains a critical challenge in advancing tomographic imaging techniques for complex biological applications.

One promising solution to these challenges is the incorporation of machine learning techniques. With the rapid advancement of machine learning, data-driven approaches have been widely and actively applied in the field of cellular biophysics, particularly for tasks such as RBC morphology classification. Numerous studies have developed reliable surrogate models based on deep learning to identify RBC shape categories using two-dimensional (2D) images~\cite{xu2017deep,simionato2021red,routt2023deep,ma2025automatic}, significantly enhancing the efficiency of diagnosis and prediction of hematological disorders. These successes in 2D classification demonstrate the potential of machine learning to effectively extract morphological features from complex biological data. For 3D geometry reconstruction of cells, traditional diffraction tomography algorithms initially relied on optimization methods to facilitate 3D phase imaging~\cite{Fienup:82,Tian:15}. Kamilov~\emph{et al.}~\cite{Kamilov:15} were among the first to introduce machine learning into this context, employing the beam propagation method (BPM) and error backpropagation to approximate the RI distribution and thereby improve 3D phase imaging. Building on this, Waller and Tian~\cite{waller2015machine} proposed the use of artificial neural networks as surrogate models to predict cellular RI distributions and reconstruct 3D geometries. Subsequent research has introduced increasingly sophisticated approaches to improve accuracy and efficiency. Waibel~\emph{et al.}~\cite{WAIBEL2022105298} proposed SHAPR by minimizing binary cross-entropy and dice loss between the true and the predicted 3D RBC shape, and later developed DISPR by incorporating a diffusion model~\cite{10230752}. Nadimpalli~\emph{et al.} further enhanced geometric reasoning by designing a neural network guided by topology-based loss~\cite{nadimpalli2023euler}. More recently, Kim~\emph{et al.}~\cite{kim2025single} developed physics-informed, coordinate-based neural networks capable of reconstructing 3D morphology directly from a single input image, without requiring angular scanning or multiple phase shifts. Despite these advances, several challenges remain unresolved. A key challenge lies in predicting the 3D geometry of RBCs from limited observational data. While incorporating physical or geometric priors as loss functions~\cite{waibel2022capturing} can improve model performance, it is not universally effective due to the substantial morphological and biomechanical variability among different RBC categories~\cite{li2017computational,xu2018stiffness,wen2025stomatocyte}, making it difficult to define unified physical constraints. Another challenge involves reducing the cost of surrogate model development, including both the computational burden of training and the effort required to prepare high-quality training datasets. While the former may be addressed through neural network architecture optimization~\cite{10230752,kim2025single} or high-performance computing resources, the latter remains a fundamental bottleneck. Training a generalizable model capable of reconstructing the 3D geometry of RBCs across diverse categories requires extensive and accurate datasets, whose preparation via conventional imaging methods is inherently time-consuming and labor-intensive.

In response to these challenges, we propose to employ a multi-fidelity neural network (MFNN)~\cite{MENG2020109020mfnn} as a promising solution. Unlike conventional surrogate neural network models that directly approximate the mapping from partial 2D data to 3D RBC geometry, MFNN leverages both high-fidelity information and prior training experience from low-fidelity approximations to enhance prediction accuracy. In the MFNN architecture, a partial 2D observation of the target RBC (regarded as high-fidelity data) is combined with a reference RBC exhibiting morphological similarity to the target RBC (regarded as low-fidelity data) to reconstruct the full 3D geometry of the target cell. This framework allows the large and costly training datasets in conventional NN models to be reduced to a set of representative samples, thereby significantly decreasing the data preparation burden discussed earlier. Motivated by the MFNN framework and our previous work on RBC morphology~\cite{wen2025stomatocyte}, we develop a surrogate predictor designed to reconstruct the 3D geometry of RBCs from partial cross-sectional 2D data. Specifically, we first introduce the architecture of the proposed predictor, which is based on MFNN and incorporates physical constraints. We then outline the theoretical foundation supporting the architecture and describe the simulation setups used to generate the training dataset. Next, we present a set of benchmark results to validate the effectiveness of the proposed model, including the positive effect of the imposed constraints in increasing the robustness of the model. Following validation, we analyze the model’s prediction accuracy and examine several key factors influencing its performance, including the RBC shape category, the geometric discrepancy between the reference and target RBCs, and the sampling strategies employed. Finally, we conclude by summarizing the predictor’s overall performance and its potential for broader applications in biophysical modeling.

\section{METHODS}
The objective of this study is to develop a surrogate predictor capable of accurately reconstructing the 3D geometric configuration of an individual RBC from conventional 2D observational data. This task represents a typical scenario of prediction from incomplete or partially observed data. To address this challenge, we employ an MFNN framework~\cite{MENG2020109020mfnn}, which utilizes the joint knowledge from both sparse high-fidelity and correlated low-fidelity sources. In our context, the prior RBC 3D shape archive obtained by experimental means of high accuracy but low efficiency can be seen as low-fidelity data, which only shares morphological similarity with the target individual RBC, but provides complete 3D geometric representations. The 2D information of the target RBC, captured through real-time imaging, is regarded as high-fidelity data, which reflects the true morphology of the target individual RBC, but contains only partial geometric information. By leveraging the complementary characteristics of both data sources (completeness from low-fidelity data and specificity from high-fidelity data), the MFNN framework enables a reconstruction of individualized 3D RBC geometries. This section provides a detailed justification for the suitability of MFNN to our problem on both theoretical and numerical foundations, introduces numerical simulation used to generate the training dataset, and outlines the predictor model framework based on MFNN and the training procedure.

\subsection{Homeomorphism Theory}
In multi-fidelity modeling theory, the low-fidelity data provide valuable trend information to bridge the gaps between sparse high-fidelity samples. By fusing heterogeneous data sources, multi-fidelity models in general can achieve significantly higher prediction accuracy based on a small set of high-fidelity data than conventional single-fidelity approaches~\cite{mfnn-doi:10.1098/rspa.2016.0751}. This relies on a fundamental assumption that a natural cross-correlation must exist between low- and high-fidelity data. The existing correlation among cell morphology can be justified mathematically using Homeomorphism theory~\cite{homeo-MOORE2007333}, which provides a rigorous topological foundation to examine the existence of a continuous and invertible mapping between two surfaces. Considering a randomly deformed 2D enclosed RBC surface, which is mathematically expressed as
\begin{equation}
    \mathcal{R}\{\mathbf{x}\} \subset\mathbb{R}^3,
\end{equation}
where $\mathbb{R}^3$ is the 3D Euclidean space, $\mathcal{R}\{\mathbf{x}\}$ can be explicit or implicit according to RBC categories. $\mathcal{R}\{\mathbf{x}\}$ of discocyte or biconcave RBC can be explicitly formulated as
\begin{equation}
    \mathcal{R}\{\mathbf{x}\} = \{(x,y,z)\in\mathbb{R}^3, z = r\sqrt{\text{1}-\frac{x^\text{2}+y^\text{2}}{r^\text{2}}}\left[c_\text{0}+c_\text{1}\frac{x^\text{2}+y^\text{2}}{r^\text{2}}+c_\text{2}\frac{{(x^\text{2}+y^\text{2})}^\text{2}}{r^\text{4}} \right]\},
\end{equation}
where ($r, c_\text{0}, c_\text{1}, c_\text{2}$) are shape parameters and empirically measured as (3.91 $\mu$m, 0.0135805, 1.001279, -0.561381)~\cite{EVANS-EA1979}. $\mathcal{R}\{\mathbf{x}\}$ of RBCs like echinocyte, acanthocyte, sphero-echinocyte, and echino-acanthocyte~\cite{Bessis-10.1007/978-3-642-88062-9_1} are implicit. Similarly, a spherical surface is mathematically expressed as
\begin{equation}
    \mathcal{S}\{\mathbf{x}\} =  \{(x,y,z)\in\mathbb{R}^3,  x^2+y^2+z^2=\text{Const}\}.
\end{equation}
Since $\mathcal{R}\{\mathbf{x}\}$ and $\mathcal{S}\{\mathbf{x}\}$ share the same topological invariants such as compactness (both surfaces are bounded and closed), connectedness (both surfaces are in one piece), and 0-genus (no holes are on both surfaces), a continuous mapping $\mathcal{F}$ naturally exists between them. The mapping $\mathcal{F}$ represents a coordinate transformation from $\mathcal{S}\{\mathbf{x}\}$ to $\mathcal{R}\{\mathbf{x}\}$ or in the inverse way. Therefore, the cross-correlation validated by Homeomorphism ensures the applicability of multi-fidelity modeling in building the surrogate predictor.

\subsection{Mesoscale simulation of SDE Transformations}
\begin{figure}[hbt!]
\centering
\includegraphics[width=1.0\linewidth]{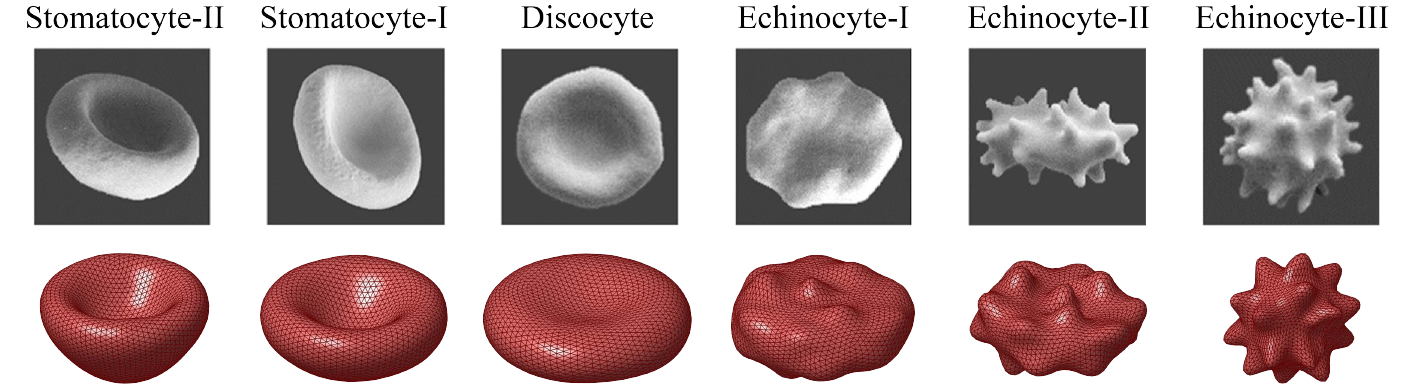}
\caption{SDE representatives' images (Stomatocyte-II, Stomatocyte-I, Discocyte, Echinocyte-I, Echinocyte-II, Echinocyte-III) observed by SEM (the first images row) and their numerical correspondence obtained by the improved two-component RBC model (the second images row). The SEM images are from the experimental reference~\cite{geekiyanage2019coarse} with authorization.}
\label{fig:M2-SDEs}
\end{figure}

Upon the theoretical validation of the mapping between the sphere and the RBC surface, a further practical validation by numerical simulation is needed. Meanwhile, our expected surrogate predictor based on MFNN is data-driven, such that it requires a reliable and easily accessible data source. Thus, we utilized an improved two-component RBC meso-model~\cite{wen2025stomatocyte} for further validation and data generation. The model has been proven capable of generating stomatocyte-discocyte-echinocyte (SDE) morphology transformations~\cite{wen2025stomatocyte} and simulating dynamical SDE blood flow in the capillary vessel~\cite{10.1063/5.0260445} in our previous research. The improved two-component RBC model incorporates the latest evidence-based biophysical characteristics and is constructed using two layers of triangular networks representing the outer lipid bilayer and the inner cytoskeleton~\cite{annurev:/content/journals/10.1146/annurev.ph.49.030187.001321}, which are connected by linking proteins between the two layers. The simulation process conducted under the framework of dissipative particle dynamics (DPD)~\cite{Li2016DPD} mimicking the mapping essentially involves energy minimization through cell membrane relaxation. During the process, RBC transforms from an initial sphere state to a stable cell membrane configuration with minimized total potential energy:
\begin{equation}
    U_{\text{total}}=U_\text{s}+(U_\text{b}+U^\text{ske}_\text{b})+(U_{\text{a+v}}+U^\text{ske}_{\text{a+v}})+U_{\text{int}},
\end{equation}
where $U_\text{s}, (U_\text{b}+U^\text{ske}_\text{b}), (U_{\text{a+v}}+U^\text{ske}_{\text{a+v}}), U_\text{int}$ denote the different potentials attributed to different mechanical behaviors of membrane stretching, bending, area- and volume-preserving, and linking protein elongation~\cite{annurev:/content/journals/10.1146/annurev.ph.49.030187.001321} (detailed model configuration and simulating setups are described in~\ref{app:A}). Our previous result~\cite{wen2025stomatocyte} indicates that a complete SDE RBC shape sequence can be obtained by modulating several mechanical and geometric parameters, which physically validates the existence of the mapping. Representatives of stable RBC shape obtained by DPD simulation and comparison of images from scanning electron microscopy (SEM) are shown in Fig.~\ref{fig:M2-SDEs}.

The dataset for training the surrogate model is constructed with multiple simulation cases of varying input parameters, and all the generated RBC shapes are within the SDE range. In each simulation case, two layers of triangular meshes, each consisting of 2562 vertices, are utilized in modeling. Since our previous research~\cite{wen2025stomatocyte} and many experimental findings~\cite{Li2015Vesi10-PhysRevE.92.012715,xu2018stiffness} indicate that the lipid bilayer and the cytoskeleton present different morphological patterns, especially in echinocyte configuration where two components detach and cytoskeleton only represents inner framework of RBC membrane, we only use 2562 vertices of bilayer representing outer morphology of RBC membrane as the training and test data in our predictor. The 3D coordinates of the initial sphere and stable RBC vertices are combined to form the training data for the surrogate predictor (configurations including DPD simulation parameters and training data formation of all the RBC cases used to train the predictor are provided in~\ref{app:A}).

\subsection{3D Predictor based on MFNN}
The key to MFNN's good approximation with incomplete data is the exploitation of the true relationship between the low- and high-fidelity data~\cite{mf2023-10.3934/acse.2023015}. A general correlation between the low- and high-fidelity data can be constructed as
\begin{equation}
    y_\text{H}=\mathcal{F}_\text{l}(x,y_\text{L})+\mathcal{F}_\text{nl}(x,y_\text{L}),
\end{equation}
where $y_\text{H}(x)$ and $y_\text{L}(x)$ are respectively the mappings in high- and low-fidelity data. $\mathcal{F}_\text{l}$ and $\mathcal{F}_\text{nl}$ respectively denote the linear and nonlinear terms in the correlation, and they together form the unknown function that maps the low-fidelity data to the high-fidelity level.

The architecture of MFNN is conventionally composed of three essential parts: the low-fidelity network $\mathcal{N}\mathcal{N}_\text{L}$ approximating the low-fidelity data, the high-fidelity network $\mathcal{N}\mathcal{N}_{\text{H}_\text{l}}$ approximating the linear correlation for the low- and high-fidelity data, and the high-fidelity network $\mathcal{N}\mathcal{N}_{\text{H}_\text{nl}}$ approximating the nonlinear correlation for the low- and high-fidelity data. The evaluation of the approximation of a neural network is known as the loss function, which is aimed to be minimized through the training process. The loss function of MFNN is expressed as
\begin{equation}
    loss_\text{total}=loss_{y_\text{L}}+loss_{y_\text{H}}+\lambda \sum \beta_\text{i}^\text{2},
\end{equation}
where $loss_{y_\text{L}}$ and $loss_{y_\text{H}}$ respectively evaluate the approximation of $\mathcal{N}\mathcal{N}_\text{L}$ and $\mathcal{N}\mathcal{N}_\text{H}$, and $\lambda \sum \beta_\text{i}^\text{2}$ denotes the $L_\text{2}$ regularization, which has been widely adopted to prevent overfitting~\cite{ZHANG2019108850}, with any weight $\beta$ in $\mathcal{N}\mathcal{N}_\text{L}$ and $\mathcal{N}\mathcal{N}_\text{H}$ and regularization rate $\lambda$.

Our predictor, based on the MFNN framework, is constructed using different kinds of neural networks. $\mathcal{N}\mathcal{N}_\text{H}$ is constructed using two forward neural networks (FNNs) representing $\mathcal{N}\mathcal{N}_{\text{H}_\text{l}}$ and $\mathcal{N}\mathcal{N}_{\text{H}_\text{nl}}$, respectively. As a fundamental type of neural network architecture, each FNN consists of multiple layers of fully connected neurons, where each layer performs an affine transformation followed by a nonlinear activation. A typical $L$-layer FNN can be written as:
\begin{equation}
    \mathscr{F}(x) = u^{(L)}\big(\cdots \sigma\big(u^{(3)}\big(\sigma\big(u^{(2)}(\sigma u^{(1)}(x))\big)\big)\big)\big),
\end{equation}
with $u^{(l)} = W^{(l)}x + b^{(l)}$, where $W$ and $b$ are the trainable weights and biases, and $\sigma$ is the activation function. In this work, we use the hyperbolic tangent (Tanh) activation for all FNN layers in $\mathcal{N}\mathcal{N}_\text{H}$.

$\mathcal{N}\mathcal{N}_\text{L}$ adopts a convolutional neural network (CNN) architecture to extract spatially localized features from input data. CNNs utilize shared-weight convolutional filters that slide across input features, producing translation-equivariant responses. A one-dimensional convolutional layer operates as:
\begin{equation}
    c=\phi(f\  *\ x+b ),
\end{equation}
where $f$ is the convolution kernel, $*$ denotes the convolution operation, $x$ is the input, $b$ is the bias, and $\phi$ is the activation function. In this work, we use the leaky rectified linear unit (LeakyReLU) activation for the CNN in $\mathcal{N}\mathcal{N}_\text{L}$. The $i$-th element of the convolution output is given by:
\begin{equation}
    (f\  *\ x)[i]=\sum^N_{j=1}x[j]\cdot f[i-j],
\end{equation}
where $N$ is the input length. To reduce dimensionality and retain dominant features, a max pooling operation follows the convolution layers. This operation selects the maximum value within a local window:
\begin{equation}
    y[i]=\text{max}\{ c[j] : j\in n\},
\end{equation}
where $n = [i \cdot s, i \cdot s + 1, \ldots, i \cdot s + w - 1]$, with stride $s$ and window size $w$. The pooled features are flattened and passed into fully connected layers, forming the output of the CNN. The entire CNN computation in $\mathcal{N}\mathcal{N}_\text{L}$ can be expressed as:
\begin{equation}
    \mathscr{C}(x) = \mathscr{F}(y(x)),
\end{equation}
where $y(x)$ represents the pooled feature map, and $\mathscr{F}$ is the previously mentioned FNN.

\begin{figure}[hbt!]
\centering
\includegraphics[width=0.85\linewidth]{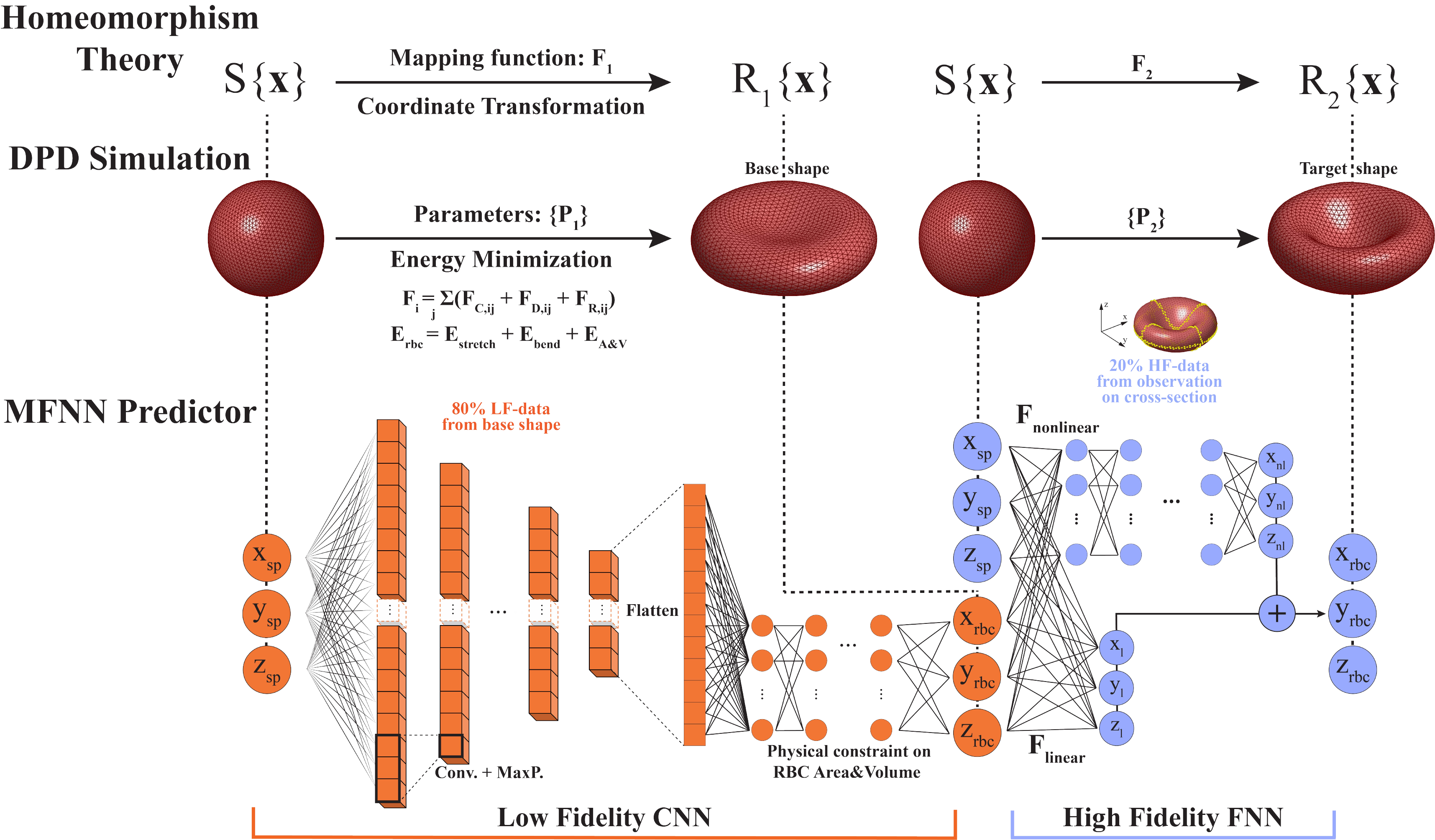}
\caption{Illustrative sketch of the identical process demonstrated by Homeomorphism theory, DPD simulation, and MFNN predictor}
\label{fig:M1-methods}
\end{figure}

As demonstrated in Fig.~\ref{fig:M1-methods}, the working process of our surrogate predictor can be divided into two major components: low-fidelity CNN and high-fidelity FNN. Low-fidelity CNN represents the training of a known reference shape or base shape with sufficient training data, where sphere coordinates $\mathbf{x}^\text{LF}_\text{sp}$ are the input, and base RBC shape coordinates $\mathbf{x}^\text{LF}_\text{rbc}$ are the output. This sphere-to-RBC deformation is essentially a coordinate transformation with strong spatial structure. Consequently, the inductive biases of convolution make the CNN a more suitable surrogate than the FNN~\cite{10.1145/3306346.3322959}. We therefore adopt the CNN as the direct surrogate for both the homeomorphic mapping and the DPD-based deformation process. High-fidelity FNN represents the approximation of the morphological correlation between base RBC shape and target RBC shape, where a joint input of base RBC shape coordinates $\mathbf{x}^\text{LF}_\text{rbc}$ from low-fidelity CNN and sphere coordinates $\mathbf{x}^\text{HF}_\text{sp}$ goes in, and the output of predicted target RBC shape coordinates $\mathbf{x}^\text{LF}_\text{rbc}$ goes out. In contrast to the low-fidelity CNN using sufficient data, the high-fidelity FNN employs an incomplete dataset containing partial observations from 2-D cross-sections for training. Overall, a prediction for the target RBC shape is generated with partial observation data of the target RBC and reference information of the base RBC. 

Informed by numerous studies demonstrating that incorporating physical laws into surrogate models can enhance their predictive capability~\cite{raissi2019physics,10.1002/smtd.202400620}, we introduce physical constraints during the training process of the base shape predictor. Among the various physical properties influencing RBC morphology, surface area and volume are particularly critical. In the low-fidelity CNN branch, the surface area and volume of the base RBC shape are known a priori and are explicitly incorporated as physical constraints during training. Therefore, the total loss to be minimized in our surrogate predictor is expressed as

\begin{align}
    loss_\text{total}&=loss_{L}+w_\text{1}*loss_{H}+w_\text{2}*loss_\text{A+V}+\lambda \sum \beta_\text{i}^\text{2},\\
    loss_{L}&=\frac{\text{{1}}}{N_{\mathbf{x}_\text{L}}}\sum^{N_{\mathbf{x}_\text{L}}}_{i=\text{1}}\left( |\mathbf{x}^\text{pred}_\text{L}-\mathbf{x}_\text{L}|^\text{2}\right),\\
    loss_{H}&=\frac{\text{{1}}}{N_{\mathbf{x}_\text{H}}}\sum^{N_{\mathbf{x}_\text{H}}}_{i=\text{1}}\left( |\mathbf{x}^\text{pred}_\text{H}-\mathbf{x}_\text{H}|^\text{2}\right),\\
    loss_\text{A+V}&=\frac{(A_\text{L}^\text{pred}-A_\text{L})^\text{2}}{A_\text{L}}+\frac{(V_\text{L}^\text{pred}-V_\text{L})^\text{2}}{V_\text{L}},
\end{align}
where $N_{\mathbf{x}_\text{L}}$ and $N_{\mathbf{x}_\text{H}}$ are respectively the sample numbers of the training data in the low-fidelity CNN and the high-fidelity FNN, subscript $(*)^\text{pred}$ represents the quantities predicted by the neural network, $A_\text{L}$ and $V_\text{L}$ are respectively the total surface area and volume of base RBC shape, and $w_\text{1}$ and $w_\text{2}$ are adjustable weights to adjust the proportion of three kinds of loss. The mathematical form of all loss functions is set as the conventional mean-square error (MSE). We will illustrate and discuss the influence of area-volume constraint on the predictor performance in Section~\ref{sec:results}. In contrast, the high-fidelity FNN branch assumes the target RBC shape to be unknown. Therefore, surface area and volume serve as key indicators for assessing the accuracy of the predicted geometry. 

\begin{figure}[hbt!]
\centering
\includegraphics[width=0.7\linewidth]{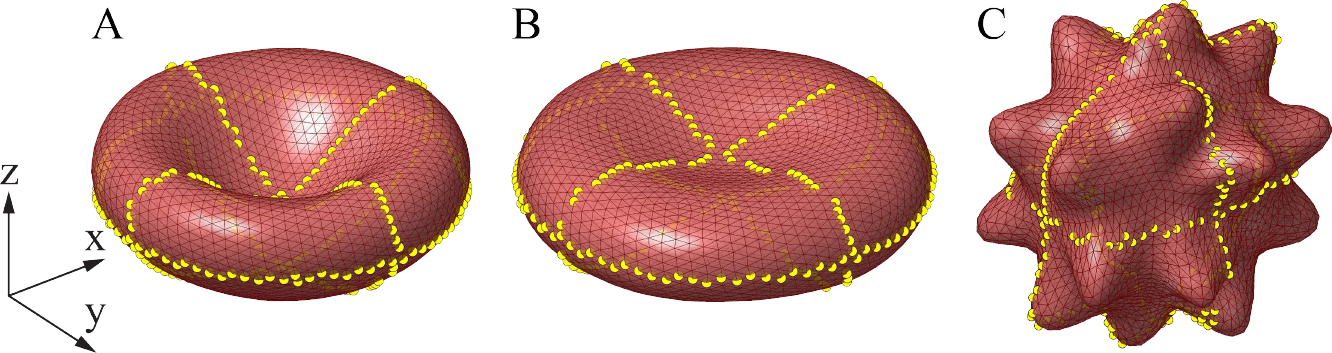}
\caption{Illustrative vertices selection of training set in MFNN: A. Stomatocyte-I, $\delta = \text{0.1}\mu m,N_\text{train}=\text{242}$; B. Discocyte, $\delta = \text{0.2}\mu m,N_\text{train}=\text{231}$; C. Echinocyte-III, $\delta = \text{0.1}\mu m,N_\text{train}=\text{263}$. Yellow dots are the highlighted selected samples on three orthogonal cross-sections with a small thickness of 2$\delta$, $N_\text{train}$ is the total number of selected vertices.}
\label{fig:M3-crossdemo}
\end{figure}

The dataset for training is composed of two parts: the base RBC set and the target RBC set, which respectively correspond to the low-fidelity CNN and the high-fidelity FNN. The base RBC set utilizes 2,000 vertices of one RBC case from the DPD simulation as the training set and the remaining 562 vertices as the test set by random sampling (R.S.), since a low-fidelity CNN represents a direct approximation of morphological transformation with sufficient known data. In contrast, a cross-sectional sampling is used to form the dataset, where the target RBC only uses 100 to 400 cross-sectionally dependent vertices as the training set, and the rest of the vertices above 2,000 as the test set. As shown in Fig.~\ref{fig:M3-crossdemo}, the target RBC training set is formed by vertices located in three orthogonal cross-sections with small thickness 2$\delta$ ($\delta<<R_\text{c}$, where $R_\text{c}$ is the characteristic radius of RBC) in Cartesian space. The small size and cross-sectional dependence of the training set are set to mimic the limited observation at several fixed angles. We will illustrate and discuss the impact of the selection and size of the training set in Section~\ref{sec:results}. In addition, to account for the rotational alignment between the target RBC and the base RBC, we apply a simple procedure that optimizes a rotation matrix to minimize the discrepancy between the two shapes (See~\ref{app:B}).

\section{RESULTS}\label{sec:results}
We consider a dataset comprising a variety of RBC shapes within the SDE range, generated through numerical simulation, as the data source to drive our surrogate model. The results are presented in two parts: the benchmark performance of the low-fidelity CNN and the prediction results of the full MFNN-based surrogate predictor. We first validate the low-fidelity CNN using six representative SDE shapes, justify the feasibility of the proposed model framework, and define the criterion used to evaluate the predictor's performance. Then, we present and discuss the prediction results of the complete MFNN-based surrogate model across all shape cases.

\subsection{Benchmark: Validation of Low-Fidelity CNN}
It is emphasized that MFNN utilizes prior experience in the mapping trained by the low-fidelity CNN, which can be independently trained outside the framework of our predictor. Therefore, validating the performance of this low-fidelity CNN is crucial and serves as a benchmark for evaluating the surrogate model. In this section, we present the mapping results of the low-fidelity CNN for six representative RBC shapes: Stomatocyte-II (Sto.II), Stomatocyte-I (Sto.I), Discocyte (Dis.), Echinocyte-I (Ech.I), Echinocyte-II (Ech.II), and Echinocyte-III (Ech.III). We also conduct an error analysis with respect to training dataset size, investigate the effect of incorporating the area-volume constraint on noise reduction during low-fidelity CNN training, compare the results of training data selected using random sampling versus cross-sectional sampling, and define the criterion to evaluate the predictor's performance.

\subsubsection{Size of training dataset}
\begin{figure}[hbt!]
\centering
\includegraphics[width=0.8\linewidth]{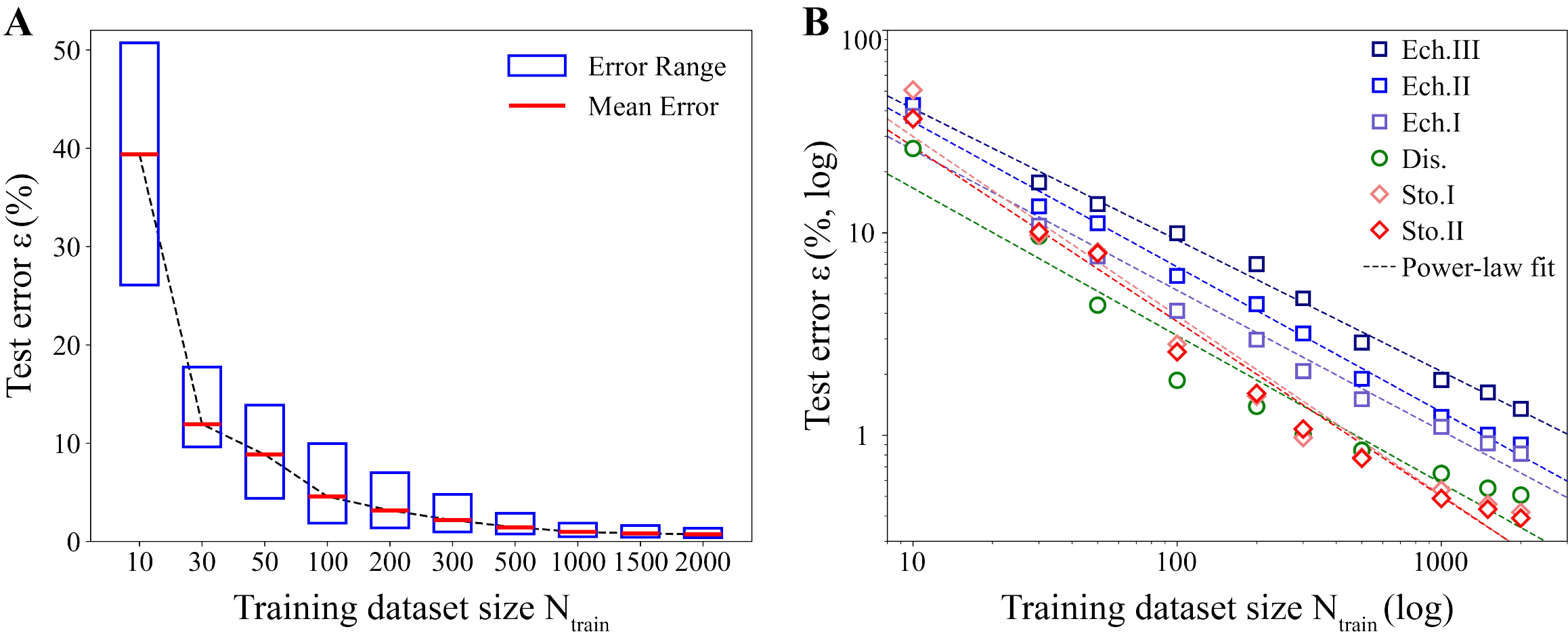}
\caption{Test error $\varepsilon$ versus training data size $N_\text{train}$ in low-fidelity CNN illustrated in A. error bar and B. two-log axis (scaling exponents and minimum test error at $N_\text{train}=\text{2,000}$ for each representative shape are Sto.II: $\alpha=\text{0.8634}$, Sto.I: $\alpha=\text{0.8854}$, Dis.: $\alpha=\text{0.7284}$, Ech.I: $\alpha=\text{0.6931}$, Ech.II: $\alpha=\text{0.7163}$, and Ech.III: $\alpha=\text{0.6488}$).}
\label{fig:R-bm1-error-size}
\end{figure}

The performance of the low-fidelity CNN and the surrogate predictor is quantitatively evaluated in terms of morphological correspondence, which is assessed with the test dataset based on coordinate alignment between the prediction shape and the reference shape. Therefore, we define accuracy $\chi$ or relative error $\varepsilon$ using $L_\text{2}$ norm,
\begin{align}
    \chi&=\left(\text{1}-\frac{ \| \mathbf{x}^{\text{pred}} - \mathbf{x}^{\text{ref}} \|_\text{2} }{ \| \mathbf{x}^{\text{ref}} \|_\text{2} }\right),\\
    \varepsilon &= \frac{ \| \mathbf{x}^{\text{pred}} - \mathbf{x}^{\text{ref}} \|_\text{2} }{ \| \mathbf{x}^{\text{ref}} \|_\text{2} },
\end{align}
where $\mathbf{x}^{\text{pred}}$ and $\mathbf{x}^{\text{ref}}$ represent the prediction coordinates and reference coordinates. Fig.~\ref {fig:R-bm1-error-size}A shows that the test error decreases as the training dataset size increases, and the mean relative error of six shapes is reduced from 39.388\% to 0.730\% when $N_\text{train}$ is increased from 10 to 2,000. The error discrepancy among the six shapes also decreases as $N_\text{train}$ increases. When $N_\text{train} = \text{2,000}$, the errors for all cases are reduced to the range of 0.390\% to 1.352\%. It is worth noting that the training data inherently contains noise, which arises from both numerical fluctuations in the DPD method~\cite{Li2016DPD} and physical fluctuations in the RBC model~\cite{peng2013lipid}. This implies the existence of a lower bound on the minimal error achievable by the surrogate model. To quantify this limit, a dimensionless standard deviation $\sigma_\text{dpd}$ representing system fluctuation of the numerical simulation is defined as
\begin{equation}
    \sigma_\text{dpd} =\frac{\sqrt{ \frac{1}{N_\text{v}} \sum_{i=1}^{N_\text{v}} \left( d_i - d_\text{mean} \right)^2 }}{R_\text{c}},
\end{equation}
where $N_\text{v}$, $d_i$, $d_\text{mean}$, and $R_\text{c}$ are the total vertex number of lipid bilayer triangular network in RBC model, the vertex distance from RBC centroid, the mean vertex distance, and the characteristic radius of RBC, respectively. As shown in Table~\ref{table:R-bm2-fluc-error}, the overall magnitude of $\varepsilon$ is consistent with $\sigma_\text{dpd}$, indicating that the low-fidelity CNN is approaching the data quality limit at $N_\text{train} = \text{2,000}$. The values of $\varepsilon$ for the echinocyte cases exhibit a larger deviation from this limit compared to the stomatocyte and discocyte cases, suggesting that the mapping task in the echinocyte scenario is slightly more difficult to approximate. For further investigation, an empirical power-law relationship for error and dataset size is referenced~\cite{hestness2017deep}. The relation is empirically summarized as $\varepsilon \propto(N_\text{train})^{-\alpha}$, by applying log on both sides, the relation is expressed as
\begin{equation}
    \text{log}(\varepsilon) = -\alpha *\text{log}(N_\text{train})+C
\end{equation}
where C denotes an arbitrary constant and $\alpha<\text{1}$ is the scaling exponent denoting a sub-linear relation between error and dataset size, a larger $\alpha$ indicates a simpler training task or a more highly structured dataset. Fig.~\ref{fig:R-bm1-error-size}B shows the result of the power-law fit, or linear regression, on a log-log scale mapping the relationship between error and dataset size. A larger $\alpha$ value is observed in the stomatocyte scenario, while a smaller $\alpha$ is seen in the echinocyte scenario. This indicates that the geometric mapping of the stomatocyte is more easily captured by the low-fidelity CNN compared to that of the echinocyte. The variation in $\alpha$ across the six shapes corresponds well with the error gap observed in Table~\ref{table:R-bm2-fluc-error}. Both patterns can be attributed to inherent morphological differences among the SDE shapes: stomatocytes and discocytes are geometrically regular, whereas echinocytes are geometrically irregular.

\begin{table}[hbt!]
\renewcommand\arraystretch{1.5}
\centering
\caption{Comparison between low-fidelity CNN prediction error $\varepsilon$ at $N_\text{train}=\text{2,000}$ and dimensionless standard deviation $\sigma_\text{dpd}$ of numerical simulation among six RBC representatives}
\label{tab:widgets}
\centering
\begin{tabular}{c|c|c|c|c|c|c}
\hline\hline
\  & Sto.II & Sto.I & Dis. & Ech.I & Ech.II & Ech.III\\ \hline
$\varepsilon$ (\%)  & 0.390 & 0.417 & 0.508 & 0.812 & 0.901 & 1.352\\ \hline
$\sigma_\text{dpd}$ (\%)  & 0.207 & 0.127 & 0.138 & 0.183 & 0.237 & 0.263\\ \hline
\hline
\end{tabular}
\label{table:R-bm2-fluc-error}
\end{table}

\begin{figure}[hbt!]
\centering
\includegraphics[width=0.85\linewidth]{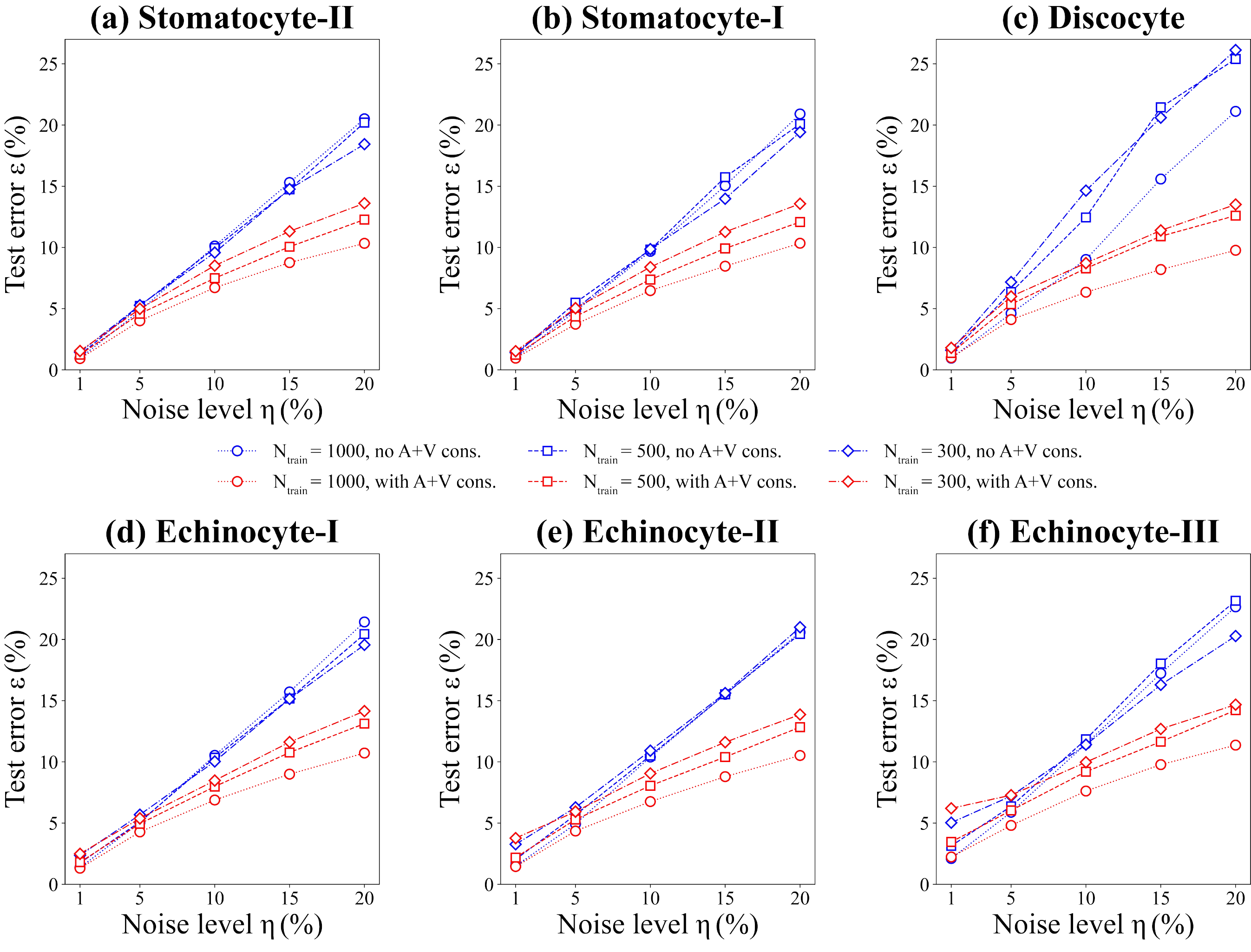}
\caption{Test error $\varepsilon$ versus inserted noise level $\eta$ in low-fidelity CNN illustrated in scenarios of six shapes: (a).Sto.II, (b).Sto.I, (c).Dis., (d).Ech.I, (e).Ech.II, and (f).Ech.III. In each scenario, cases of varying $\eta=\text{1\%},\text{5\%},\text{10\%},\text{15\%},\text{20\%}$ and $N_\text{train}=\text{300},\text{500},\text{1,000}$ are given for comparison.}
\label{fig:R-bm3-error-noise}
\end{figure}

\subsubsection{Role of area-volume constraint}
The purpose of introducing the area-volume physical constraint is to enhance the model's performance by incorporating physical information during training. Contrary to our expectations, the introduction of the area-volume constraint does not reduce prediction error in the ``clean data'' scenario, where the dataset contains only DPD-induced noise. However, a noise-reduction effect emerges when the area-volume constraint is applied in scenarios with imposed additional noise. We present a comparison of the low-fidelity CNN's performance with and without the physical constraint across six representative RBC shapes, under varying conditions of $N_\text{train}$ and $\eta$, where $\eta$ represents the noise level. A Gaussian white noise is added to the output coordinate data, and the mathematical process is interpreted as
\begin{equation}
    \tilde{\mathbf{x}}_\text{rbc} = \mathbf{x}_\text{rbc}\left(\text{1}+\eta\ \xi\right), \label{eqn:noise}
\end{equation}
where $\xi$ is a Gaussian random variable with zero mean and unit variance, and $\eta \in[\text{0.01, 0.2}]$ is an adjustable parameter setting the noise level. As shown in Fig.~\ref{fig:R-bm3-error-noise}, at a low added noise level of $\eta = \text{1\%}$, where the data can be considered approximately clean, the prediction error $\varepsilon$ is slightly higher in all cases when the area-volume constraint is imposed. This suggests that, under clean conditions, the constraint worsens the prediction performance. We found that the training loss declines more slowly when the area-volume constraint is applied, indicating that in the "clean data" scenario, the constraint acts not as a guide but rather as a barrier during loss minimization, leading to degraded performance. However, this negative effect vanishes when the noise level $\eta$ exceeds 5\%. At higher noise levels, the area-volume constraint exhibits a positive noise-reducing effect. As illustrated in Fig.~\ref{fig:R-bm3-error-noise}, all six shape cases show a substantial reduction in error when the constraint is applied. Specifically, at $\eta = \text{20\%}$, the prediction error decreases by 26.19\% to 53.75\%. This reduction across all cases demonstrates that imposing the area-volume constraint enhances the robustness of the low-fidelity CNN, making it essential for maintaining the stability of our MFNN predictor. Furthermore, the contrasting effects of the area-volume constraint in clean versus noisy data scenarios imply that surface area and volume are valid but not decisive parameters in regulating RBC morphology. It is also worth noting that the discocyte shows the highest sensitivity to increasing noise and the greatest error reduction from the imposed constraint, as seen in Fig.~\ref{fig:R-bm3-error-noise}(c). This indicates that discocyte morphology is more easily disrupted by noise and more difficult for the neural network to capture under such conditions. This observation aligns with our previous finding that the discocyte, being naturally selected for its biconcave shape, appears uniquely within a narrow range of biomechanical conditions~\cite{wen2025stomatocyte}.

\subsubsection{Random sampling versus cross-sectional sampling}
\begin{figure}[hbt!]
\centering
\includegraphics[width=0.9\linewidth]{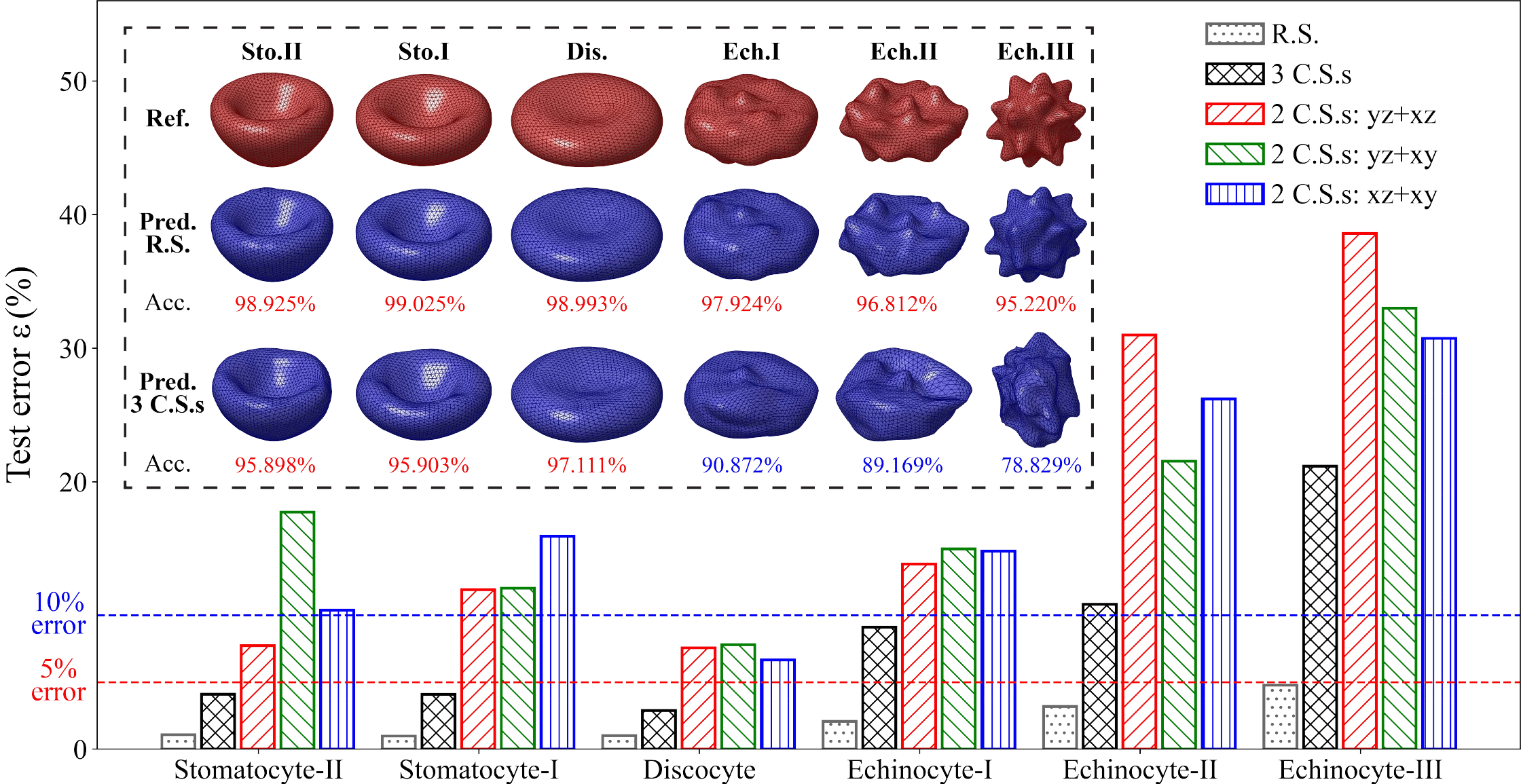}
\caption{The comparison of test error $\varepsilon$ among random samplings (R.S.) and cross-sectional samplings of 4 kinds: 3 cross-sections (3 C.S.s) and 2 cross-sections (2 C.S.s) in combination of x-y cross-section, x-z cross-section, and y-z cross-section. The inner subplots of RBCs illustrate the corresponding predicted morphology and accuracy with R.S. data and 3 C.S.s data. All the cases of R.S. for the comparison are set $N_\text{train}=\text{300}$, and the cases of C.S. sampling are set $N_\text{train}=\text{165}\sim\text{258}$, depending on the selection of cross-sections.}
\label{fig:R-bm4-error-3d2d}
\end{figure}

The simple low-fidelity CNN has demonstrated reliable predictive performance using a small size of training dataset with random sampling. For example, the average accuracy across the six representative shapes reaches 97.81\% when $N_\text{train} = 300$. However, random sampling is impractical in real experimental settings, as data obtained through conventional methods is typically in the form of 2D images~\cite{li2013continuum}. Therefore, it is crucial to emphasize that a practical surrogate predictor should be capable of using partial information from specific cross-sectional observations to reconstruct the full 3D geometry of RBCs. As previously mentioned, we use cross-sectional sampling to mimic the partial data from an experimental source, and each cross-section is defined as an orthogonal plane with small thickness $\delta$. Thus, we tested the performance of the low-fidelity CNN with 4 types of cross-sectional sampling schemes:
\begin{itemize}
    \item Three cross-sections (3 C.S.s): x–y, x–z, and y–z cross-sections.
    \item Two cross-sections (2 C.S.s): x–y and x–z cross-sections.
    \item Two cross-sections (2 C.S.s): x–y and y–z cross-sections.
    \item Two cross-sections (2 C.S.s): x–z and y–z cross-sections.
\end{itemize}

As demonstrated in Fig.~\ref{fig:R-bm4-error-3d2d}, the overall prediction error of cross-sectional sampling is notably higher than that of random sampling across all six shapes. Cases using 2 C.S.s consistently perform worse than those using 3 C.S.s. This discrepancy can be naturally attributed to the differing levels of geometric information present in the datasets. Although the number of samples is comparable across datasets formed by different sampling strategies, random sampling selects vertices distributed throughout the global surface of the shape, resulting in a more geometrically informative dataset. In contrast, cross-sectional sampling restricts vertex selection to local orthogonal planes, thereby providing limited morphological information and leading to reduced prediction accuracy. Moreover, the geometric information level continues to degrade as the reduced dimensions in cross-sectional sampling. In the inner subplots of Fig.~\ref{fig:R-bm4-error-3d2d}, we demonstrate the direct morphological comparison among three kinds of samplings. The predicted RBCs more closely resemble the reference when the prediction accuracy satisfies $\chi \geq \text{95\%}$, where the overall morphology and local geometry highly resemble the original reference, indicating an accurate prediction. As $\chi$ decreases into the range of $\text{90\%} \sim \text{95\%}$, several mismatches appear in local geometry, indicating a moderate prediction. Multiple mismatches and general morphological discrepancies start to appear as $\chi$ drops below 90\%, indicating a poor prediction. When $\chi$ falls below 85\%, a completely different geometry appears, which can be seen as a failure in shape prediction. The mentioned relationship between prediction accuracy and visual morphological resemblance can be used as the criterion to evaluate the predictor's performance in the following analysis. Using this criterion, only three cases in the discocyte and stomatocyte scenarios with 3 C.S.s training data demonstrate accurate prediction, and a few cases in the discocyte and stomatocyte scenarios with 2 C.S.s training data demonstrate moderate prediction. In contrast, most cases using cross-sectional data, especially in the echinocyte scenario, are considered to fail in providing effective predictions. As indicated in the previous analysis, stomatocyte and discocyte are geometrically regular, and their morphology can be more easily informed by cross-sectional data, while the echinocyte's irregular morphology is harder to capture through cross-sectional sampling.

To conclude, we tested the performance of the single low-fidelity CNN under varying dataset sizes, justified the role of the area-volume constraint in enhancing the network's robustness, and defined the criterion used to evaluate the predictor's performance. Although predictions relying solely on the low-fidelity CNN with partial cross-sectional data are not fully reliable, this evaluation provides both validation of the methodology and valuable insights for setting neural network parameters in the subsequent surrogate predictor based on the MFNN framework.

\subsection{Prediction of Multi-Fidelity Neural Network}
As described above, the surrogate predictor reconstructs the morphology of the target RBC by learning two key components: the transformation mapping from the base shape, and the correlation between the base and target shapes. These correspond to the low-fidelity CNN and high-fidelity FNN components of the model, respectively. Following the validation of the low-fidelity CNN, this section presents the performance of the MFNN in predicting target RBC geometries and investigates several influencing factors, including the category of SDE, the geometric discrepancy between the base shape and the target RBC, and the strategies used for cross-sectional sampling in the echinocyte scenario. An analysis evaluating the robustness of the predictor is also presented.

\subsubsection{MFNN Prediction across SDE}
In multi-fidelity modeling theory, the correlation between the low-fidelity and high-fidelity regimes is regarded as a crucial source of additional information that enhances the performance of the approximation. In our case, this correlation reflects the morphological similarity between the base RBC and the target RBC, which is expected to be captured by the linear and non-linear components of the high-fidelity FNN. Morphological similarity plays a key role in ensuring prediction accuracy, reinforcing the principle that the base shape used in the low-fidelity CNN should geometrically resemble the target shape. Following this principle, we select a base shape from the same geometric category as the target, based on the classic RBC classification by Bessis~\cite{Bessis-10.1007/978-3-642-88062-9_1}. In our study, we introduce a slight modification to the conventional classification by dividing all SDE shapes into two broad categories: stomato-discocyte and echinocyte. This categorization is informed by our results from the low-fidelity CNN, which indicate that stomatocytes and discocytes are geometrically regular and relatively easy to approximate, while echinocytes are geometrically irregular and more difficult to approximate. Under these two broad categories, we further define five specific shape types: Stomatocyte, Discocyte, Echinocyte-I, Echinocyte-II, and Echinocyte-III. For illustration of cases, we define case label as \textit{X.Y.(n)}, where \textit{X} and \textit{Y} are respectively the specific shape of the base shape and the target shape, and \textit{(n)} is the specific case number. If \textit{X} is equal to \textit{Y}, we consider that a prediction is conducted within a single type, while an across-type prediction is conducted if \textit{X} is not equal to \textit{Y}. Additionally, it is essential to acknowledge that the selection process relies on prior knowledge of the target shape category, which appears to necessitate artificial experience. However, the determination of target shape category can be replaced by a surrogate model, numerous studies have introduced a variety of surrogate models identifying RBC shape category with decent accuracy~\cite{xu2017deep,simionato2021red,routt2023deep,ma2025automatic}. In our study, we manually choose the base shape for each target shape. Since we assume the base shape is priorly accessible, $N_\text{train,L}=\text{2,000}$ is assigned for the low-fidelity CNN, and $N_\text{train,H}$ of around $\text{250}$, depending on cross-sectional sampling in each case, is assigned for high-fidelity FNN. The detailed setups for all MFNN cases are given in~\ref{app:C}.

Fig.~\ref{fig:R-accdissmap}A presents a morphology map of MFNN cases, plotted along two axes: prediction accuracy $\chi$ and a dimensionless dissimilarity metric $\mathcal{D}^\text{*}$, which quantifies the discrepancy between the base shape (or low-fidelity shape) and the target shape (or high-fidelity shape). The dimensionless dissimilarity is defined as
\begin{align}
  \mathcal{D}^\text{*} &= \frac{d_\text{chamfer}}{R_\text{c}},\\
    d_\text{chamfer} &=\sqrt{ \frac{\text{1}}{|P|} \sum_{\mathbf{p} \in P} \min_{\mathbf{q} \in Q} \| \mathbf{p} - \mathbf{q} \|_\text{2} + \frac{1}{|Q|} \sum_{\mathbf{q} \in Q} \min_{\mathbf{p} \in P} \| \mathbf{q} - \mathbf{p} \|_\text{2}},
\end{align}  
where $d_\text{chamfer}$ is the root chamfer distance, which is commonly used to evaluate the difference between a point cluster $P = \{ \mathbf{p}_\text{1}, \mathbf{p}_\text{2}, \dots, \mathbf{p}_n \} \subset \mathbb{R}^d$ and a point cluster $Q = \{ \mathbf{q}_\text{1}, \mathbf{q}_\text{2}, \dots, \mathbf{q}_m \} \subset \mathbb{R}^d$ in machine learning~\cite{5539837}, $P$ and $Q$ respectively refer to base shape and target shape in our content. As demonstrated in Fig.~\ref{fig:R-accdissmap}A, the predictions for most cases using 3 C.S.s and 2 C.S.s datasets with the MFNN achieve an accuracy above 95\%. In contrast, the previous single low-fidelity CNN produced an average accuracy below 85\% on the same datasets, indicating that the advanced structure of the MFNN improves model performance. However, predictions based on only a 1 C.S. dataset largely failed to reconstruct the target shape, suggesting that at least two cross-sectional observations are necessary to provide sufficient information for the surrogate predictor. In terms of category in RBC shape type, consistent with our previous result of low-fidelity CNN, stomato-discocyte group cases exhibit higher accuracy than those of echinocyte due to a different level of regularity in geometry. Furthermore, the prediction accuracy $\chi$ decreases as the discrepancy, quantified by $\mathcal{D}^\text{*}$, between base shape and target shape increases. In both groups of stomato-discocyte and echinocyte cases, the across-type case yields a higher error than the single-type ones, which is attributed to the increased $\mathcal{D}^\text{*}$ due to the gap of different RBC shape types. We further demonstrated a more quantified result in Fig.~\ref{fig:R-accdissmap}B, where a linear regression is applied to display the trend of $\chi$ as $\mathcal{D}^\text{*}$ grows. A steeply declining trend line, fitted to the cases with a 1 C.S. dataset, indicates poor performance in MFNN prediction. In contrast, the more gradual trend lines fitted to the cases with 2 C.S.s and 3 C.S.s datasets maintain accuracy levels above 90\% until $\mathcal{D}^\text{*}$ exceeds 0.25, demonstrating the reliability of MFNN predictions under these conditions.

In addition to the general trend of the morphology map, it is interesting to perform analysis in the specific case of the map. In the discoycte cases (\textit{D.D.(1)} and \textit{D.D.(2)}), except for good predictions given with 2 C.S.s dataset and 3 C.S.s dataset, the 1 C.S. datasets of y-z and x-z plane also infer reliable predictions, while only the prediction of 1 C.S. datasets of x-y plane failed. This pattern can be attributed to the simple geometric nature of the biconcave, which is that the geometry of the biconcave can be mathematically deduced into two geometric quantities: radius of the "disk" and depth of the symmetrical cave. As illustrated in Fig.~\ref{fig:R-accdissmap}C, the "side view" information by either y-z or x-z cross-section provides both quantities, while the "top view" information by x-y cross-section only provides one, thus MFNN with 1 C.S dataset of x-y plane failed in giving accurate prediction in cave depth, resulting in the "non-cave" or "wrong-cave" geometry shown in the map. A similar analysis can be applied to the three stomatocyte cases (\textit{S.S.(1)}, \textit{S.S.(2)}, and \textit{S.S.(3)}). A "non-cave" scenario is observed in the case with a 1 C.S. dataset from the x–y plane. However, the other two cases with 1 C.S. datasets also fail to provide accurate predictions, and the variation in dissimilarity increases across the stomatocyte cases. Based on the resulting accuracy trends, an empirical order of cross-sectional informativeness is observed: x–z cross-section $>$ y–z cross-section $>$ x–y cross-section. However, the underlying cause for this ordering and the anomalies in prediction performance remains unclear. A similar uncertainty arises in the across-type case \textit{D.S.}, where extreme errors are found in predictions using 1 C.S. datasets from the x–z and y–z planes. Despite these unexplained phenomena in the \textit{S.S.} and \textit{D.S.} cases, the predictions using 2 C.S.s data remain reliable.

\subsubsection{Echinocyte scenario}
\begin{figure}[!th]
\centering
\includegraphics[width=0.9\linewidth]{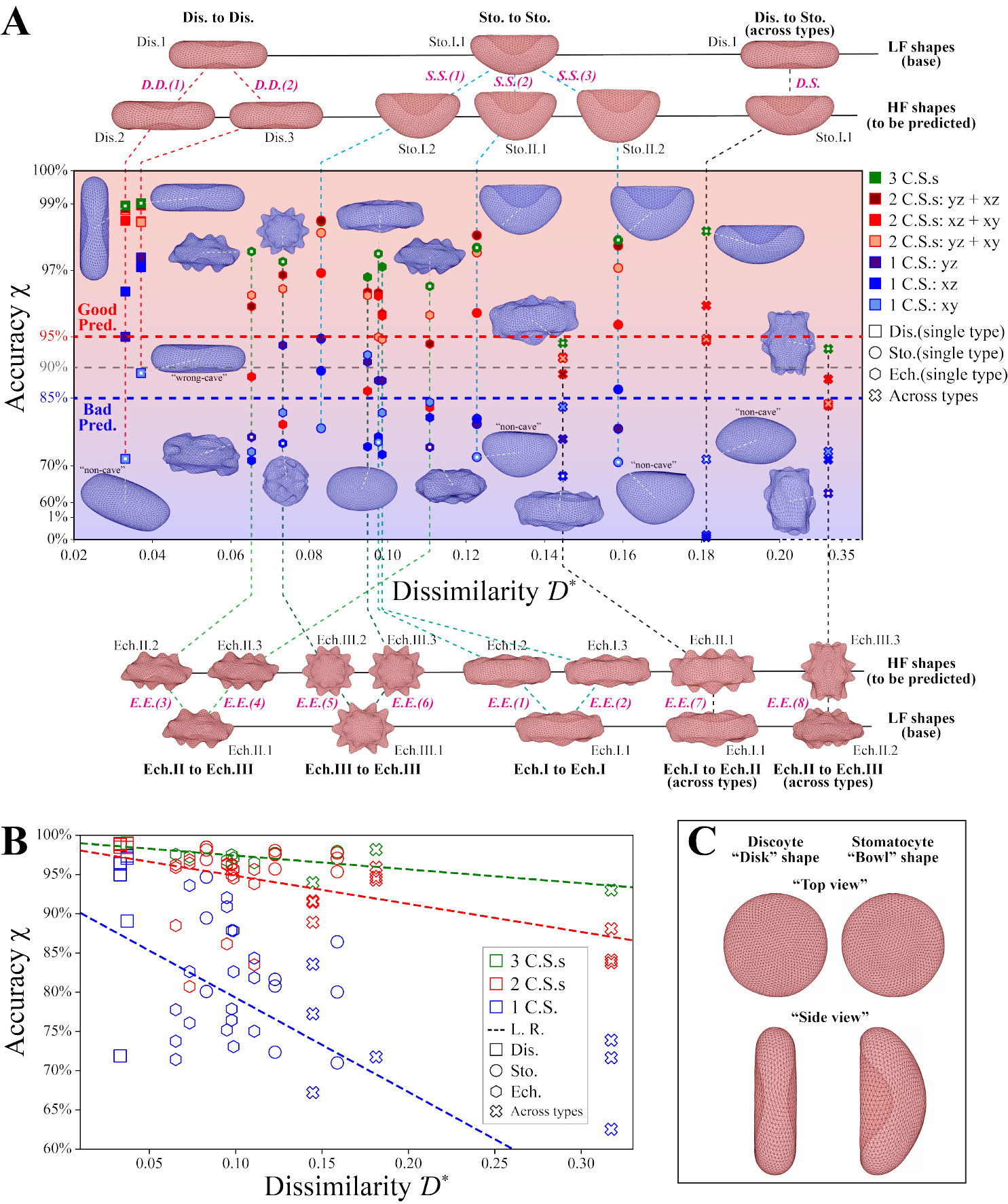}
\caption{The prediction map described by accuracy $\chi$ and dissimilarity $\mathcal{D}^\text{*}$ among all cross-sectional cases of SDE predicted by MFNN. A. $\chi$-$\mathcal{D}^\text{*}$ map where the directly predicted morphology and their corresponding reference morphology (LF and HF shapes) of major cases are presented. Each case is referred as \textit{X.Y.(n)} where \textit{X}, \textit{Y}, and \textit{(n)} respectively represent LF shape category, HF shape category, and specific case label. B. $\chi$-$\mathcal{D}^\text{*}$ relation where a linear regression (L.R.) is used for 3 C.S.s cases, 2 C.S.s cases, and 1 C.S. case, separately. C. Illustration of "top view" and "side view" of "disk"-shaped discocyte and "bowl"-shaped stomatocyte.}
\label{fig:R-accdissmap}
\end{figure}

\begin{figure}[!bh]
\centering
\includegraphics[width=0.85\linewidth]{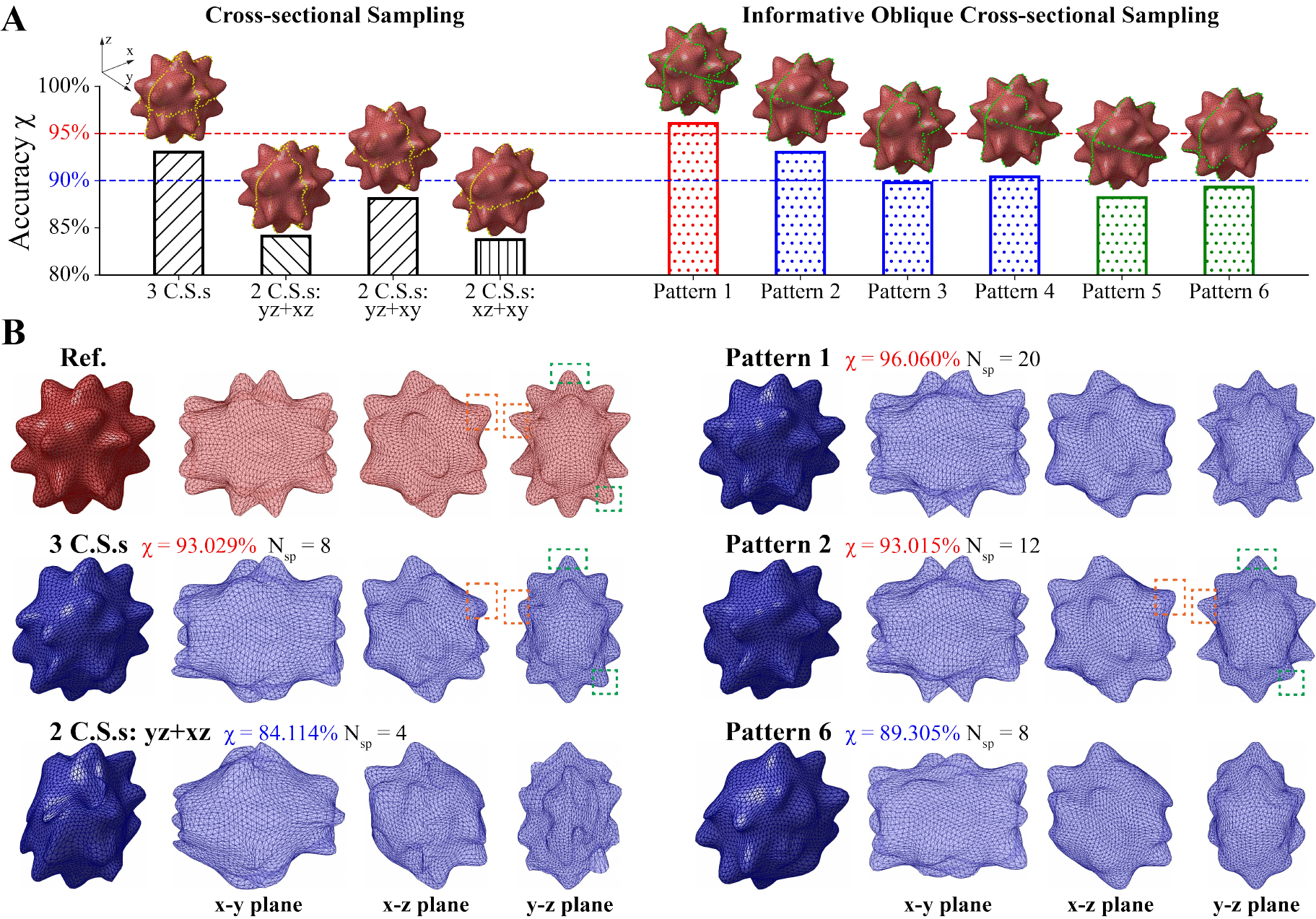}
\caption{The comparison of predictions in the case of \textit{E.E.(8)} using orthogonal cross-sectional sampling and informative oblique cross-sectional (I.O.C.) sampling. A. Prediction accuracy using a variety of cross-sectional sampling methods. B. Demonstration of morphological prediction using 3 C.S.s, 2 C.S.s: yz+xz, I.O.C. pattern 1, I.O.C. pattern 2, and I.O.C. pattern 5. $N_\text{sp}$ represents the number of spicules sampled in the training dataset, the total number of Ech.III.3 is 20. The orange and green dashed boxes highlight the discrepancy in morphology approximation between the cases of 3 C.S.s and pattern 2.}
\label{fig:R-pick}
\end{figure}

The unique irregular geometry of echinocyte cases is characterized by the presence of spicules or bumps on the RBC membrane surface~\cite{Samamtha2022Morphological}. Accurately capturing the morphology of each spicule is essential for the MFNN to make reliable predictions. As illustrated in the map of Fig.~\ref{fig:R-accdissmap}A, echinocyte cases with a "bad" prediction outcome ($\chi<\text{85\%}$) fail to approximate the spicules accurately, and the predicted shapes can hardly be recognized as echinocytes. In contrast, echinocyte cases with a "good" prediction outcome ($\chi>\text{95\%}$) exhibit both an accurate global RBC framework and precise reconstruction of local spicules on the membrane surface. We further investigate the effect of informative oblique cross-sectional (I.O.C.) sampling, which directly captures vertices located in the tip regions of the spicules (demonstrations of different patterns are in Fig.~\ref{fig:R-pick}A). Fig.~\ref{fig:R-pick}A demonstrates an overall improvement in prediction accuracy when the MFNN is trained using datasets generated through I.O.C. sampling with various patterns, as shown in the case of \textit{E.E.(8)}. We defined $N_\text{sp}$ to represent the number of sampled spicules, where the total number of spicules in Echinocyte-III is 20. With sampling in Pattern 1, five oblique cross-sections capture all 20 spicules. As a result, the MFNN yields an accurate prediction with $\chi = 96.06\%$, and all locally predicted spicules are geometrically consistent with the target Echinocyte-III, as shown in Fig.~\ref{fig:R-pick}B. The cases using 3 C.S.s and Pattern 2 demonstrate comparable prediction accuracy at $\chi \sim \text{93\%}$, as both involve sampling from three cross-sections. However, when comparing local spicule alignment (highlighted in the orange and green dashed boxes in Fig.~\ref{fig:R-pick}B), geometric discrepancies are observed in different local spicule regions between the two cases. These differences arise from the variation in information provided by the respective cross-sectional sampling datasets. A further comparison is made between the cases of 2 C.S.s: yz+xz and Pattern 6, where the datasets are both sampled from two cross-sections. The Pattern 6 case yields a more accurate prediction than the 2 C.S.s: yz+xz case in terms of overall geometry observed from the x–y plane. This improvement is attributed to the greater amount of spicule information included in the Pattern 6 dataset, even though both cases result in poor prediction accuracy ($\chi < \text{90\%}$). 

\begin{figure}[!th]
\centering
\includegraphics[width=0.8\linewidth]{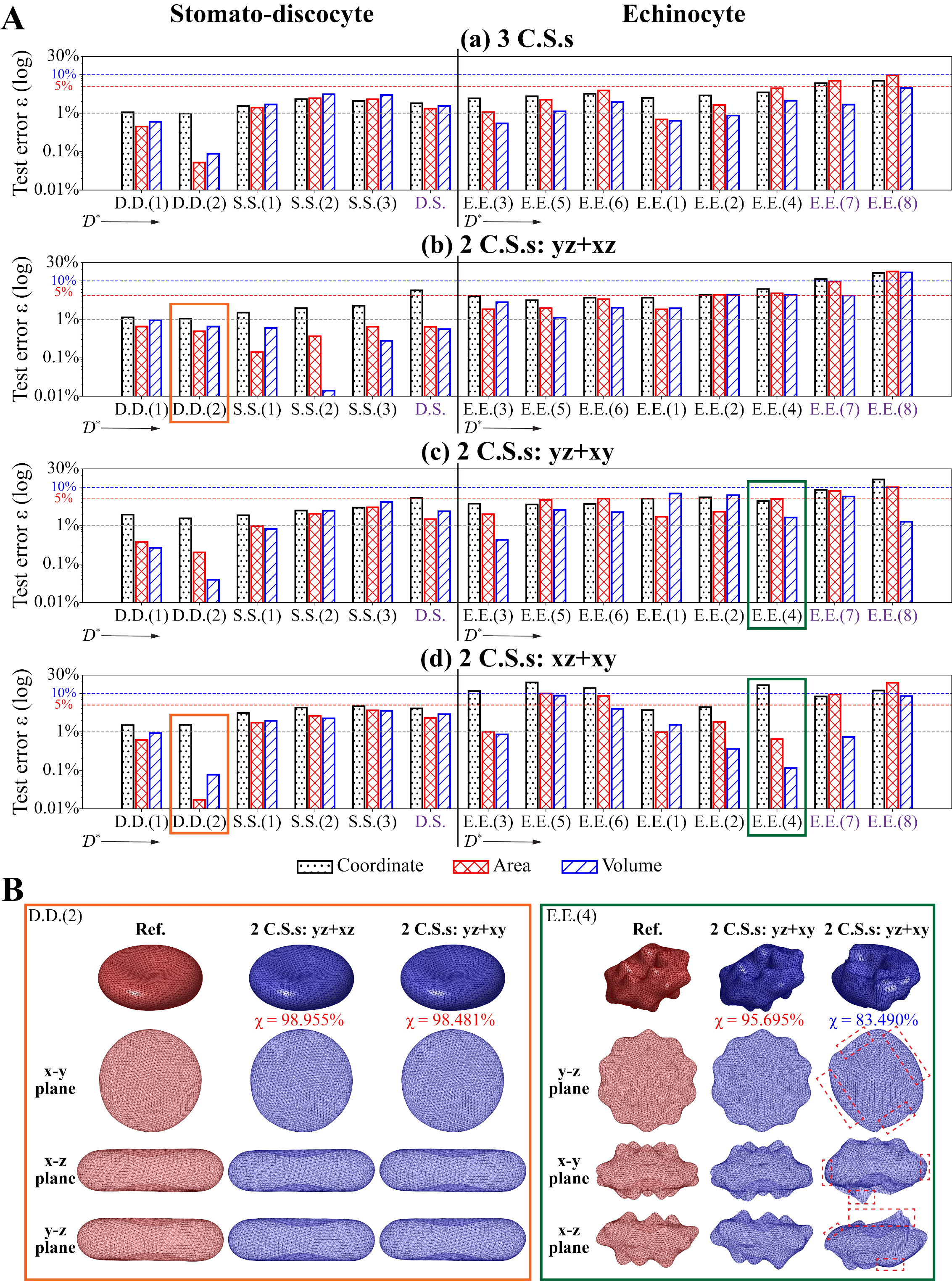}
\caption{The comparison of test error among all SDE cases of 3 C.S.s and 2 C.S.s predicted by MFNN and morphological comparison in cases of \textit{D.D.(2)} and \textit{E.E.(4)}. A. The test error on coordinates, surface area, and volume of all SDE cases of 3 C.S.s and 2 C.S.s predicted by MFNN. All the bar subplots are divided into the stomato-discocyte group and the echinocyte group, and cases are ordered in growth of $\mathcal{D}^\text{*}$. B. Demonstration of morphological prediction of \textit{D.D.(2)} and \textit{E.E.(4)} at different angles of orthogonal planes, red dashed boxes highlight major misalignment in the poor prediction case.}
\label{fig:R-accdissbar}
\end{figure}

In addition, the echinocytic pattern in Ech.II and Ech.III can be uniquely characterized by a disk-like framework with spicules distributed across the surface~\cite{Samamtha2022Morphological,wen2025stomatocyte}, where more spicules appear on the flat side of the disk (as illustrated in the example \textit{E.E.(4)} in Fig.~\ref{fig:R-accdissbar}B). As a result, cross-sectional sampling data that includes fewer vertices from the flat side is less informative with respect to spicule geometry. This explains the low prediction accuracy observed in the \textit{E.E.(3,4,5,6)} cases in Fig.~\ref{fig:R-accdissmap}A, where no y–z cross-section (representing the flat side) was included in the sampled dataset. Both the enhancement in prediction performance due to I.O.C in the case of \textit{E.E.(8)} and poor prediction results in the cases of \textit{E.E.(3,4,5,6)} highlight the critical role of spicule morphology in accurately predicting echinocyte geometry.

As we previously indicated, the area and volume of the RBC are vital and effective parameters in measuring the RBC morphology. Differing from the role as an auxiliary physical constraint to improve neural network training by reducing noise in the low-fidelity CNN, the area–volume alignment in the high-fidelity FNN serves as a crucial evaluator of MFNN performance. To evaluate the area-volume alignment in prediction, we define $\chi_\text{A}$, $\chi_\text{V}$, $\varepsilon_\text{A}$, and $\varepsilon_\text{V}$, expressed as
\begin{align}
    \chi_\text{A}&=\text{1}-\frac{|A_\text{H}^\text{pred}-A_\text{H}|}{A_\text{H}},\\
    \varepsilon_\text{A} &= \frac{|A_\text{H}^\text{pred}-A_\text{H}|}{A_\text{H}},\\
    \chi_\text{V}&=\text{1}-\frac{|V_\text{H}^\text{pred}-V_\text{H}|}{V_\text{H}},\\
    \varepsilon_\text{V} &= \frac{|V_\text{H}^\text{pred}-V_\text{H}|}{V_\text{H}},
\end{align}
where $A_\text{H}^\text{pred}$, $A_\text{H}$, $V_\text{H}^\text{pred}$, and $V_\text{H}$ are the predicted RBC surface area, the actual target RBC surface area, the predicted RBC volume, and the actual target RBC volume. Fig.~\ref{fig:R-accdissbar}A presents the values of $\varepsilon$, $\varepsilon_\text{A}$, and $\varepsilon_\text{V}$ for all SDE predictions made by the MFNN. In the majority of cases, $\varepsilon_\text{A}$ and $\varepsilon_\text{V}$ are positively correlated with $\varepsilon$, indicating that a certain level of coordinate agreement generally ensures comparable alignment in surface area and volume. However, a contradiction is observed in a few cases where the low $\varepsilon_A$ and $\varepsilon_V$ are accompanied by a high $\varepsilon$ in coordinate alignment (highlighted in orange and green boxes in Fig.~\ref{fig:R-accdissbar}A). A further elaboration on specific morphological behavior is illustrated in Fig.~\ref{fig:R-accdissbar}B. In the case of \textit{D.D.(2)}, the prediction using 2 C.S.s: xz+xy cross-sectional sampling yields lower errors in surface area and volume ($\varepsilon_\text{A} \sim \text{0.01}\%$, $\varepsilon_\text{V} \sim \text{0.1}\%$) compared to the prediction using 2 C.S.s: yz+xz sampling ($\varepsilon_\text{A} \sim \varepsilon_\text{V} \sim \text{1}\%$). Meanwhile, the coordinate error $\varepsilon$ shows the opposite trend: the $\varepsilon$ value for the yz+xz case is slightly lower than that for the xz+xy case. Both \textit{D.D.(2)} cases exhibit good approximation in morphology, and no obvious discrepancy is observed in direct morphological comparison. In addition to the discocyte cases, two cases of \textit{E.E.(4)} demonstrate a more contrasting pattern: the prediction with higher area–volume alignment using 2 C.S.s: xz+xy cross-sectional sampling results in a poorer performance in coordinate prediction, whereas the prediction with lower area–volume alignment using 2 C.S.s: yz+xy sampling yields better predictive accuracy. To be morphologically specific (highlighted in red dashed boxes in Fig.~\ref{fig:R-accdissbar}B), the former prediction yields $\chi=\text{83.49\%}$ and fails to retain echinocytic identity in the y-z plane as no visible bumps are predicted. Furthermore, mismatches appear in several local spicule regions in both the x–y and x–z plane views. These two exceptional cases indicate that achieving good area–volume alignment does not guarantee a reliable prediction result, although good coordinate consistency generally ensures good area–volume consistency. This observation aligns with our earlier analysis of the imposed area–volume constraints in the low-fidelity CNN, which suggested that while surface area and volume are essential parameters, they are not decisive in modulating RBC morphology. Therefore, parameters such as the area-to-volume ratio, which are widely recognized as vital in the biophysical processes of RBC morphological transformations~\cite{geekiyanage2019coarse}, may primarily act as resultant indicators rather than causal factors in RBC shape changes, particularly for SDE transformations. As a result, investigations focusing on the effects of microstructural activities of the RBC membrane~\cite{xu2018stiffness,wen2025stomatocyte} may be essential for understanding the underlying mechanisms of RBC morphological transformations in future research.

\begin{figure}[!ht]
\centering
\includegraphics[width=0.9\linewidth]{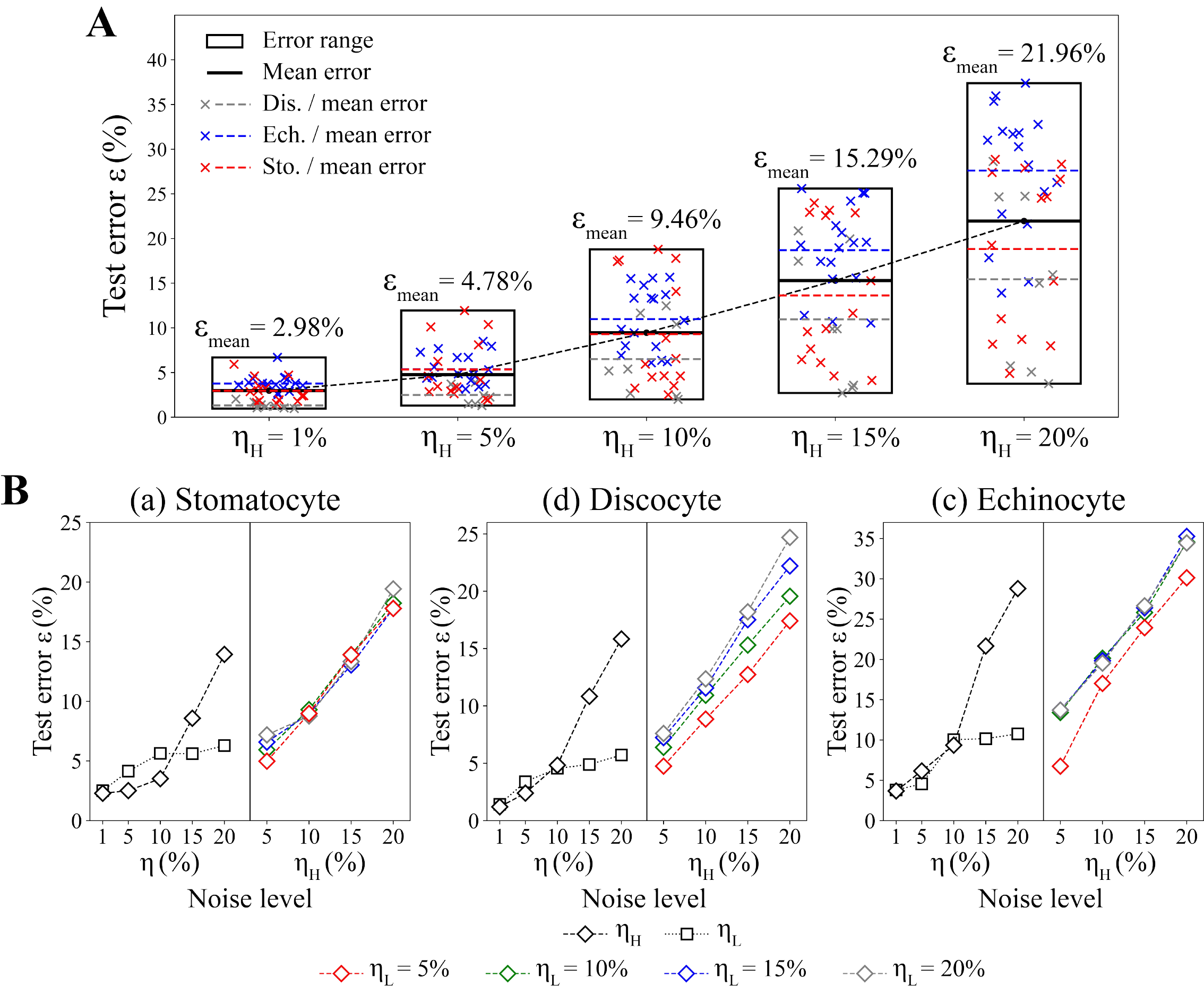}
\caption{Test error $\varepsilon$ versus inserted noise level $\eta$ of MFNN predictions. A. Error bar of $\varepsilon$ versus $\eta_\text{H}$ (only high-fidelity data is contaminated) of MFNN cases (3 C.S.s and 2 C.S.s cases of $\chi\geq \text{95\%}$ in Fig.~\ref{fig:R-accdissmap}A are used in noise analysis). B. $\varepsilon$ versus $\eta$ in three specific cases (\textit{S.S.(2)} with 2 C.S.s: yz+xz, \textit{D.D.(1)} with 2 C.S.s: yz+xz, and \textit{E.E.(3)} with 2 C.S.s: yz+xz), where three scenarios are compared: only low-fidelity data is contaminated, only high-fidelity data is contaminated, and both low- and high-fidelity data are contaminated. The right subplot represents the former two scenarios, and the left subplot represents the latter scenario.}
\label{fig:R-mfnoise}
\end{figure}

\subsubsection{Analysis of model robustness}
Understanding the robustness of a surrogate model is critical when the training dataset is affected by unavoidable noise. In the context of predicting 3D RBC geometry, both the low-fidelity data representing the reference RBC geometry and the high-fidelity data representing partial observations of the target RBC may be subject to contamination. To evaluate robustness under different conditions, we consider three scenarios: (1) only the low-fidelity data is contaminated, (2) only the high-fidelity data is contaminated, and (3) both low- and high-fidelity data are contaminated. The insertion of noise in the dataset is identically interpreted as Eq.~(\ref{eqn:noise}) as previously mentioned, and the noise level in low-fidelity data and high-fidelity data are respectively $\eta_\text{L}$ and $\eta_\text{H}$. Fig.~\ref{fig:R-mfnoise}A illustrates the overall relation between $\varepsilon$ and $\eta_\text{H}$ when only the partial cross-sectional data of the target RBC is contaminated, and cases used for the noise test in the MFNN predictor are 3 C.S.s and 2 C.S.s cases of $\chi\geq \text{95\%}$ in Fig.~\ref{fig:R-accdissmap}A. The mean coordinate error, $\varepsilon_\text{mean}$, across all tested cases exhibits a linear increase with respect to $\eta_\text{H}$, suggesting that MFNN performance degrades proportionally as noise intensity increases. This trend indicates a limited robustness of the predictor to perturbations in the high-fidelity input. 

However, the stability patterns observed in the stomato-discocyte cases and the echinocyte cases show a significant difference. In the stomato-discocyte cases, $\varepsilon$ exhibits a slow-increasing trend, with a mean accuracy of $\chi = \text{82.87\%}$ at $\eta_\text{H} = \text{20\%}$. In contrast, the echinocyte cases display a more rapid increase in $\varepsilon$, with a lower mean accuracy of $\chi = \text{72.40\%}$ at the same noise level. To further investigate, we compare three specific cases—\textit{S.S.(2)} with 2 C.S.s: yz+xz, \textit{D.D.(1)} with 2 C.S.s: yz+xz, and \textit{E.E.(3)} with 2 C.S.s: yz+xz—as shown in Fig.~\ref{fig:R-mfnoise}B. In response to all three noise scenarios, the $\varepsilon$ values for the echinocyte case are consistently higher than those for the two stomato-discocyte cases. This indicates that the training process for approximating echinocyte morphology is more sensitive to noise, and that the MFNN predictor exhibits greater instability in the presence of perturbations when modeling geometrically irregular shapes. In contrast, the MFNN predictor demonstrates robustness when approximating stomato-discocyte morphology under the scenario of sole high-fidelity noise insertion ($\eta_\text{H}$). The discrepancy in prediction performance across different RBC shapes can be naturally attributed to variations in their geometric regularity. In terms of the MFNN predictor's response to contaminated data sources, Fig.~\ref{fig:R-mfnoise}B shows that the coordinate error $\varepsilon$ exhibits opposite trends in response to $\eta_\text{L}$ and $\eta_\text{H}$ when both are increased to 10\%. Specifically, $\varepsilon$ increases rapidly with rising $\eta_\text{H}$, whereas it converges more slowly with increasing $\eta_\text{L}$, indicating that the predictor is more sensitive to noise in the high-fidelity input than in the low-fidelity input. The consistent appearance of this pattern across all three specific cases indicates that the overall stability of the MFNN is more sensitive to biased high-fidelity data than to low-fidelity data when subjected to large perturbations ($\eta_\text{H} > \text{10\%}$). The lower sensitivity to $\eta_\text{L}$ can be attributed to the area–volume constraints imposed in the low-fidelity CNN, which were previously shown to significantly enhance the neural network’s robustness. Moreover, it is interesting to note that $\eta_\text{L}$ affects $\varepsilon$ differently across the three cases when both $\eta_\text{L}$ and $\eta_\text{H}$ are introduced. In the stomatocyte case, the influence of $\eta_\text{L}$ on $\varepsilon$ becomes less apparent as $\eta_\text{H}$ increases. A similar pattern is observed in the echinocyte case when $\eta_\text{L} \geq \text{10\%}$. In contrast, the discocyte case exhibits a more pronounced increase in $\varepsilon$ as $\eta_\text{L}$ increases, regardless of the value of $\eta_\text{H}$. This behavior can be further explained by interpreting $\eta_\text{L}$ as the perturbation to the learned prior knowledge of the reference shape, which plays a more critical role in predicting discocyte morphology. Conversely, $\eta_\text{H}$ represents a perturbation to the high-fidelity corrective information used to refine the predictor’s approximation to the target shape, which is more influential in predicting stomatocyte and echinocyte morphology. To summarize, the prediction stability is more sensitive to noise from the high-fidelity data source than the low-fidelity one, since the negative effect of noise is reduced by area-volume constraints in the low-fidelity CNN. The robustness of the MFNN varies with the complexity of the predicted RBC morphology, resulting in a moderate level of overall robustness in predictor performance.

\section{Conclusion and Discussions}
Our study presents an MFNN-based surrogate model capable of accurately reconstructing 3D RBC morphology from partial 2D cross-sectional data. A key advantage of the proposed framework lies in its ability to generalize across diverse RBC morphologies without requiring case-specific algorithmic modifications. In contrast, the reconstruction quality of conventional approaches, both model-based and learning-based, is highly dependent on the underlying algorithm, which often needs to be manually adapted to extract the morphological features of different cell types. For example, many studies aimed at improving ODT imaging techniques have relied on subtle adjustments to the reconstruction algorithm to better capture the RI distribution~\cite{Sung:09, 10.1371/journal.pone.0262106}. Similarly, several recent deep learning-based studies have refined surrogate 3D reconstruction models using shape-specific biophysical constraints tailored to particular cell morphologies~\cite{WAIBEL2022105298,waibel2022capturing,nadimpalli2023euler}. While such modifications can enhance performance, they often come at the cost of generalizability, specifically, the model’s ability to handle varying geometries across RBC categories. Alternatively, the MFNN avoids this pitfall by incorporating prior structural knowledge through low-fidelity data that represent base or reference shapes. As a result, the network is not tasked with reconstructing 3D geometry from scratch using 2D inputs alone. Instead, it evolves a new 3D target shape from a morphologically similar 3D reference under the guidance of high-fidelity 2D observations. In this framework, morphological information specific to each target shape is not re-learned through additional training or algorithmic adjustment, but is instead encoded through the selection of an appropriate reference shape, thereby significantly reducing the effort required to adapt the model to new RBC geometries.

Extensive research has established successful precedents showing that incorporating externally determined physical factors as constraints can improve the performance of surrogate models~\cite{raissi2019physics,shi2024physics}. Following this rationale, we introduce the surface area and volume of RBCs, which have been long recognized as vital parameters in modulating RBC morphological transformations~\cite{geekiyanage2019coarse}, to inform the neural networks as the physical constraint. However, unlike many previous studies, our results reveal a contrasting pattern: while the imposed area–volume constraint does not enhance model performance on clean data, it significantly improves robustness under noisy conditions. On the one hand, the lack of improvement with clean data suggests that area and volume play a non-decisive role in governing RBC morphological transformations. In particular, the area-volume constant in our surrogate model is not as governing as the differentiation equations in physics-informed neural networks (PINN)~\cite{raissi2019physics}, which predicts an accurate solution using the differentiation equations and the sole samples from boundary conditions. On the other hand, the unexpected noise-reducing effect of the constraint highlights its regulatory function, maintaining the model's stability when the data source is contaminated. Moreover, the further area-volume evaluation result of MFNN, which shows a sufficient but not necessary relationship between coordinates consistency and area-volume consistency, also proves the vital but non-decisive role of area and volume in RBC morphological transformations. Although the primary focus of this work is on 3D geometry reconstruction, this finding aligns with recent biophysical studies that emphasize the role of subcellular microstructural activities within RBCs~\cite{xu2018stiffness,wen2025stomatocyte}. These studies suggest that traditionally emphasized parameters such as the area-to-volume ratio, though empirically useful as indicators, may serve more as resultant outcomes of shape transformations rather than causal drivers.

Beyond the role of physical constraints, several influencing factors were found to affect the MFNN’s predictive performance. Since our surrogate model relies on learning the morphological correlation between a base shape and a target RBC, the degree of shape resemblance quantified by dissimilarity $\mathcal{D}^*$ provides a guiding principle for pairing an appropriate reference shape for each target RBC. Specifically, the chosen reference shape should belong to the identical specific shape type as the target shape (i.e., Sto., Dis., Ech.I, Ech.II, or Ech.III). Although the full 3D geometry of the target RBC is unknown and must be predicted, its shape type must be identified in advance to guide reference selection. Currently, our pairing process relies on manual selection. However, many studies have shown that this step can be automated using deep learning models trained to classify RBC shape types from a single 2D cross-sectional image~\cite{xu2017deep,simionato2021red,routt2023deep,ma2025automatic}, indicating that the entire prediction pipeline could eventually operate without manual oversight. Another important influencing factor is the RBC category itself. Both the low-fidelity CNN benchmark and the MFNN achieve higher prediction accuracy for stomato-discocyte cases and lower accuracy for echinocyte cases. This difference can be attributed to geometric complexity: stomato-discocytes have smooth, symmetric surfaces, whereas echinocytes contain irregular spicule structures that are more difficult to reconstruct. On a biophysical level, echinocyte formation is also associated with more intricate subcellular processes, including lipid bilayer and cytoskeleton detachment~\cite{pan2018super,wen2025stomatocyte} as well as cytoskeletal rearrangement~\cite{kruger2018actin}, which introduce additional variability that further complicates prediction. The sampling strategy has a measurable impact on performance, particularly for echinocytes. Predictions using I.O.C. sampling, which captures spicule features more directly, consistently outperform those using conventional orthogonal sampling. This improvement arises because I.O.C. sampling provides more informative input regarding spicule geometry and distribution, which dominates echinocyte morphology.

Despite the promising performance of the proposed surrogate predictor in predicting 3D RBC morphology, several limitations remain. First, although the MFNN framework is structurally designed for generalizability, its current validation is confined to simulation data and a limited subset of RBC morphologies within the SDE range. This reflects a restriction in test case diversity rather than a limitation of the model itself. To more fully evaluate its potential, future work should incorporate experimentally acquired datasets and expand the morphological spectrum to include pathological or abnormal RBC types, such as spherocytes and elliptocytes in hereditary blood disorders~\cite{li2012two}, and sickle-shaped RBCs in SCA~\cite{10.1172/JCI106273}. Furthermore, in physiological conditions, RBCs continuously undergo deformation as they travel through narrow capillaries or microfluidic channels, often adopting highly irregular or transient shapes. Including such dynamic flow scenarios in future modeling would more accurately reflect in vivo conditions and expose the model to a broader range of realistic shape variations. These extensions are essential for advancing the model’s utility in biomedical applications such as disease diagnosis, morphological classification, and treatment monitoring. Second, the current model setup adopts a simplified approach to both rotational alignment and cross-sectional sampling. Rotation is handled by optimizing a single matrix to minimize the geometric difference between the input and reference shapes, which is suitable for controlled data but may fail under more variable experimental conditions. Incorporating rotation-invariant architectures could enhance the model's flexibility. Similarly, cross-sectional sampling is based on selecting point clouds within narrow bands around fixed planes, such as $\lvert x \rvert < \delta$ for y-z cross-sections. While this approximates realistic imaging setups, actual experimental cross-sections are typically acquired at precisely $x = 0$. Although this introduces only minimal error, fixed-plane sampling may miss local shape features, especially in irregular RBCs. Adaptive, feature-aware sampling strategies could better reflect real imaging practices and improve prediction accuracy. Third, the MFNN framework exhibits only moderate robustness under noisy conditions. While area–volume constraints improve the stability of the low-fidelity CNN, the overall model is more sensitive to perturbations in high-fidelity data. Prediction performance degrades more rapidly as high-fidelity noise increases, indicating a greater reliance on clean corrective input. In contrast, low-fidelity noise has a smaller effect, owing to the regularization provided by the imposed physical constraints. Robustness also varies with shape complexity: geometrically regular cells, such as stomatocytes and discocytes, maintain higher stability than irregular shapes like echinocytes. To enhance noise resilience, future work may adopt noise-aware training or regularization techniques such as Bayesian neural networks~\cite{MENG2021110361} or uncertainty-based loss functions~\cite{https://doi.org/10.1002/qre.1027}.

In this study, we developed a surrogate predictor to reconstruct the full 3D RBC geometry from 2D cross-sectional information. The established model is based on MFNN, which efficiently utilizes the joint knowledge from both sparse high-fidelity data and correlated low-fidelity data. The fundamental structure of the predictor is divided into two parts: the low-fidelity CNN and the high-fidelity FNN. Inspired by the Homeomorphism theory~\cite{homeo-MOORE2007333} and our previous numerical simulation~\cite{wen2025stomatocyte}, where RBC geometry deforms from an initial sphere to a stable configuration through energy minimization, the former neural network learns the geometric formation process of a base RBC shape, while the latter learns the morphological correlation between a base shape and a target shape. Our results demonstrate an overall prediction accuracy above 95\% in the majority of cases using sample data from at least two cross-sections. Several factors, including the dissimilarity between the target and base shapes, RBC category, and cross-sectional sampling strategy, were found to influence prediction performance. The unique structure of MFNN enables our surrogate 3D predictor to achieve reliable performance, minimize data preparation effort, and maintain strong generalization across diverse RBC morphologies. Despite current limitations in robustness, sampling design, and dataset diversity, these findings highlight the potential of MFNN-based frameworks in advancing geometry-aware cellular modeling. The proposed approach offers a scalable and data-efficient solution for 3D reconstruction tasks in complex cellular environments.

\section*{Acknowledgments}
This work was supported by the National Science Foundation via projects CDS\&E-2204011 and DGE-2244342.

\appendix

\section{Stomatocyate-Discocyte-Echinocyte Transformations Simulation}\label{app:A}
In the present study, the improved two-component RBC model is employed to simulate the quasi-static process of RBC deformation under the DPD numerical framework, thus generating a dataset for training of our surrogate predictor. Extensive numerical research demonstrated that DPD applied at the mesoscopic scale is highly effective for RBCs due to its ease of use and flexibility in modeling complex structured fluids~\cite{peng2013lipid,fedosov2010systematic,li2014probing}. The improved two-component RBC model based on DPD has been applied to investigate SDE Transformations of RBCs~\cite{wen2025stomatocyte} and their impact on apparent viscosity in dynamic capillary blood vessel~\cite{10.1063/5.0260445} in our previous work. The DPD methodology and the improved two-component RBC model can be seen in the \textit{METHOD} \& \textit{SUPPLEMENTARY MATERIAL} in the previous publication~\cite{wen2025stomatocyte}. 

The DPD simulation results utilized in the training dataset are shown as follows. By modulating mechanical and geometric parameters in the improved two-component RBC model, a variety of SDE RBCs are obtained. A rigorous analysis of the relation between those parameters and stable SDE geometry was conducted in our previous research~\cite{wen2025stomatocyte}. The simulation setups for each case representing a stable RBC shape involved as the training and test samples in our surrogate predictor are given in Table.~\ref{tab:dpd}. All the simulation cases are built with the mesh density of $N_\text{d}=16$ and computed on an AMD 7965WX 24-core CPU platform. The average computation time of one case is 27.67 minutes.

\begin{table}[htbp]
\centering
    \caption{DPD simulation setups for all SDE cases ($N_\text{c}$ and $r_\text{sh}$ are dimensionless, and the rest of the parameters are in DPD value)}
\label{tab:dpd}
\begin{tabularx}{\textwidth}{c|c c c c c c c c c|c}
\toprule
\textbf{Shape id} & $k_\text{b}$ & $\mu$ & $N_\text{c}$ & $A_\text{0}$ & $V_\text{0}$ & $r_{\text{sh}}$ & $k_\text{a}$ & $k_\text{A}$ & $k_\text{V}$ & \textbf{Shape Demo.} \\
\midrule
Dis.1 (Repre. Dis.) & 96.12 & 6.44 & 16 & 132.57 & 91.75 & 0.0 & 4,900 & 5,000 & 5,000 & \adjustbox{valign=m}{\includegraphics[width=0.12\textwidth]{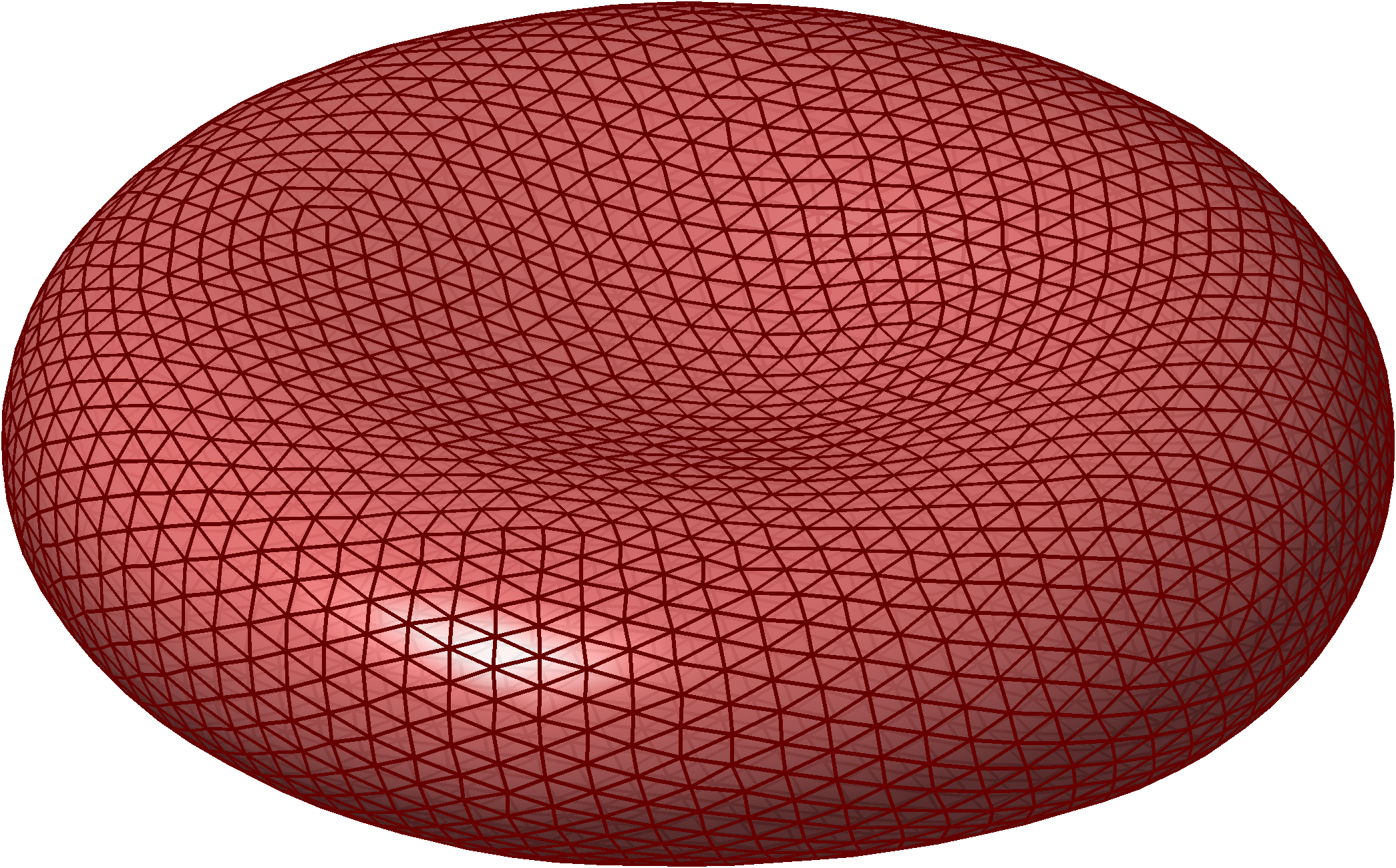}} \\[12pt]
Dis.2 & 96.12 & 6.44 & 16 & 139.20 & 91.75 & 0.0 & 4,900 & 5,000 & 5,000 & \adjustbox{valign=m}{\includegraphics[width=0.126\textwidth]{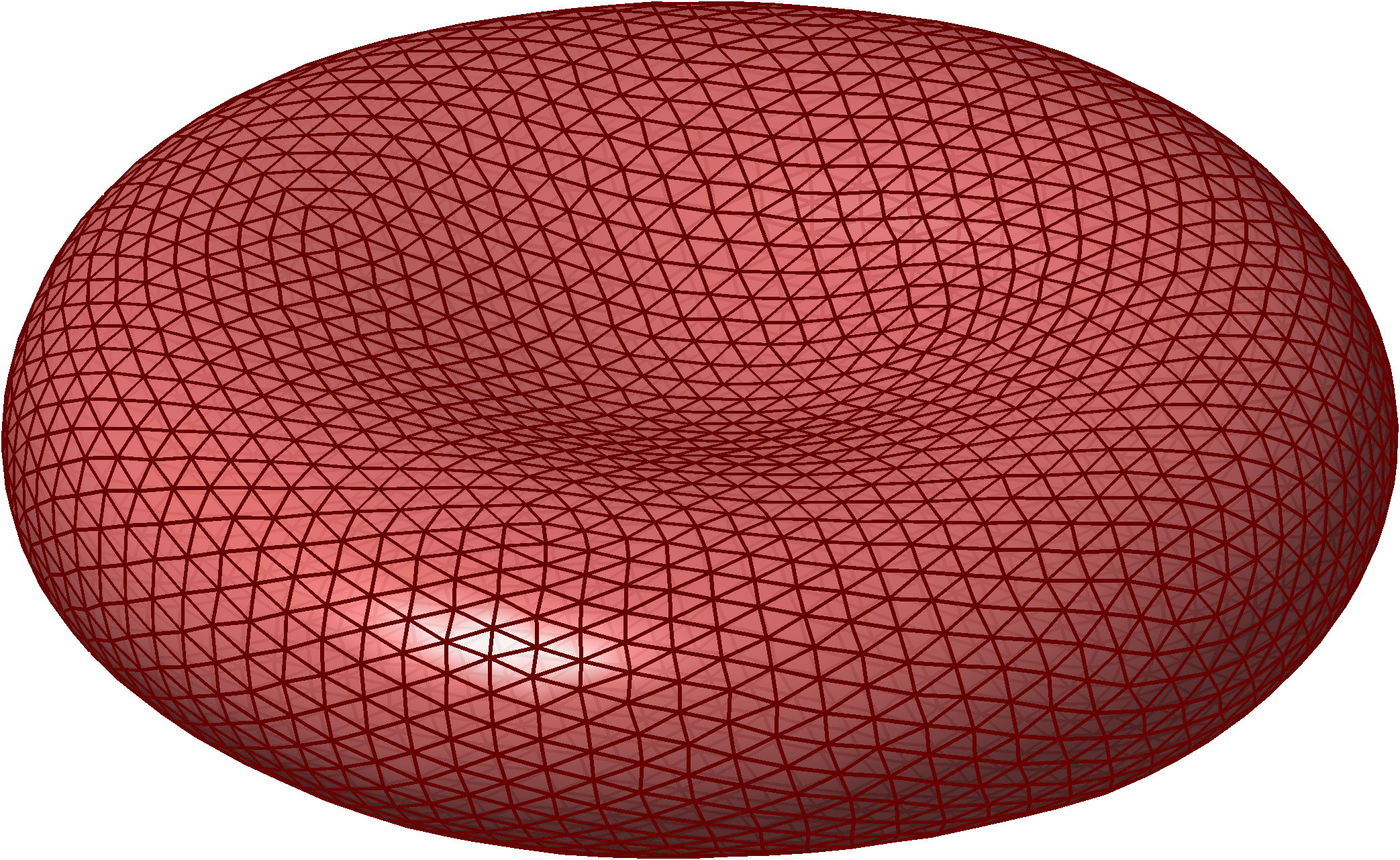}}  \\[12pt]
Dis.3 & 96.12 & 6.44 & 16 & 125.95 & 91.75 & 0.0 & 4,900 & 5,000 & 5,000 & \adjustbox{valign=m}{\includegraphics[width=0.114\textwidth]{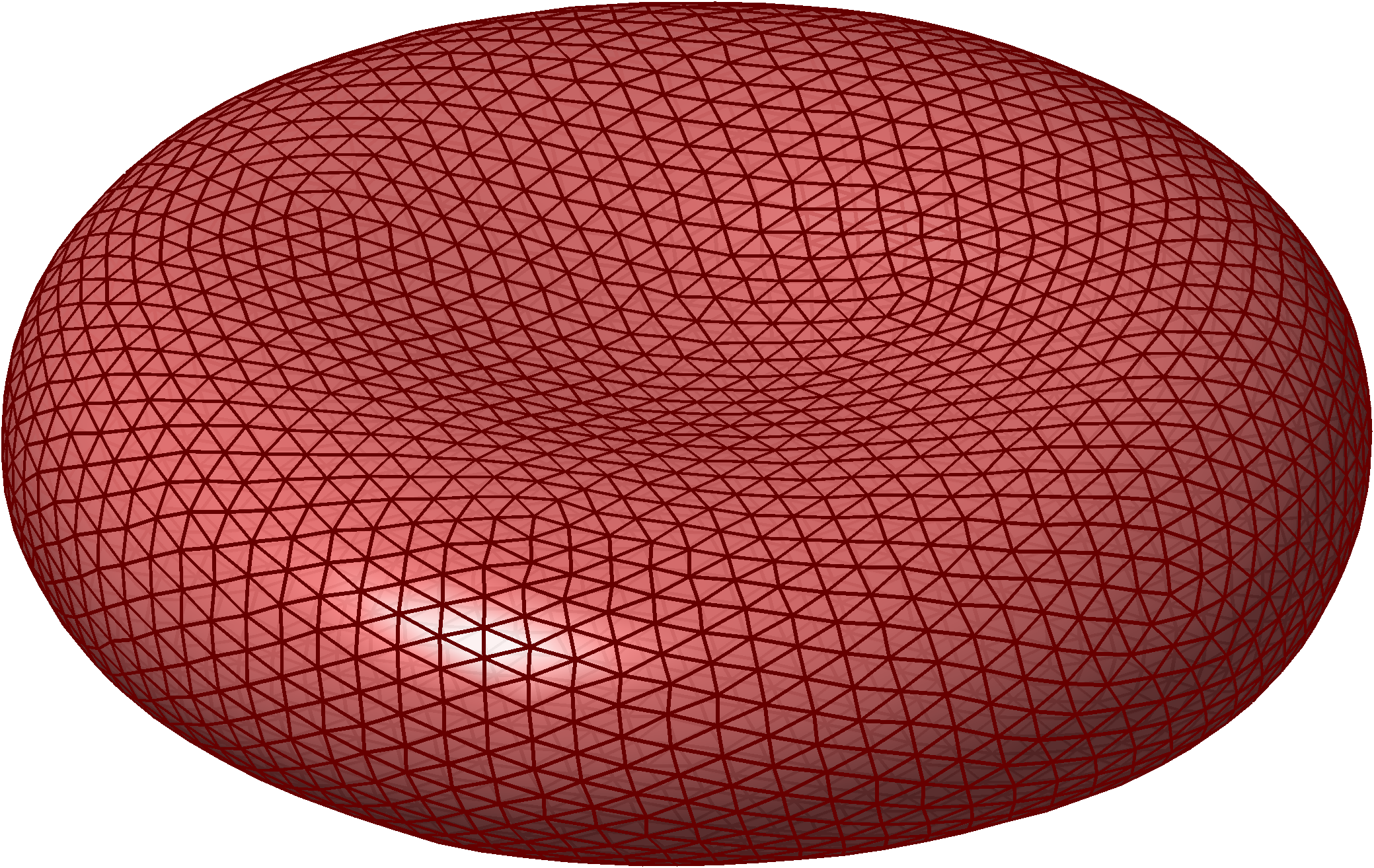}}  \\[12pt]
Sto.I.1 (Repre. Sto.I) & 96.12 & 80.0 & 16 & 132.57 & 91.75 & 0.0 & 4,900 & 5,000 & 5,000 & \adjustbox{valign=m}{\includegraphics[width=0.1\textwidth]{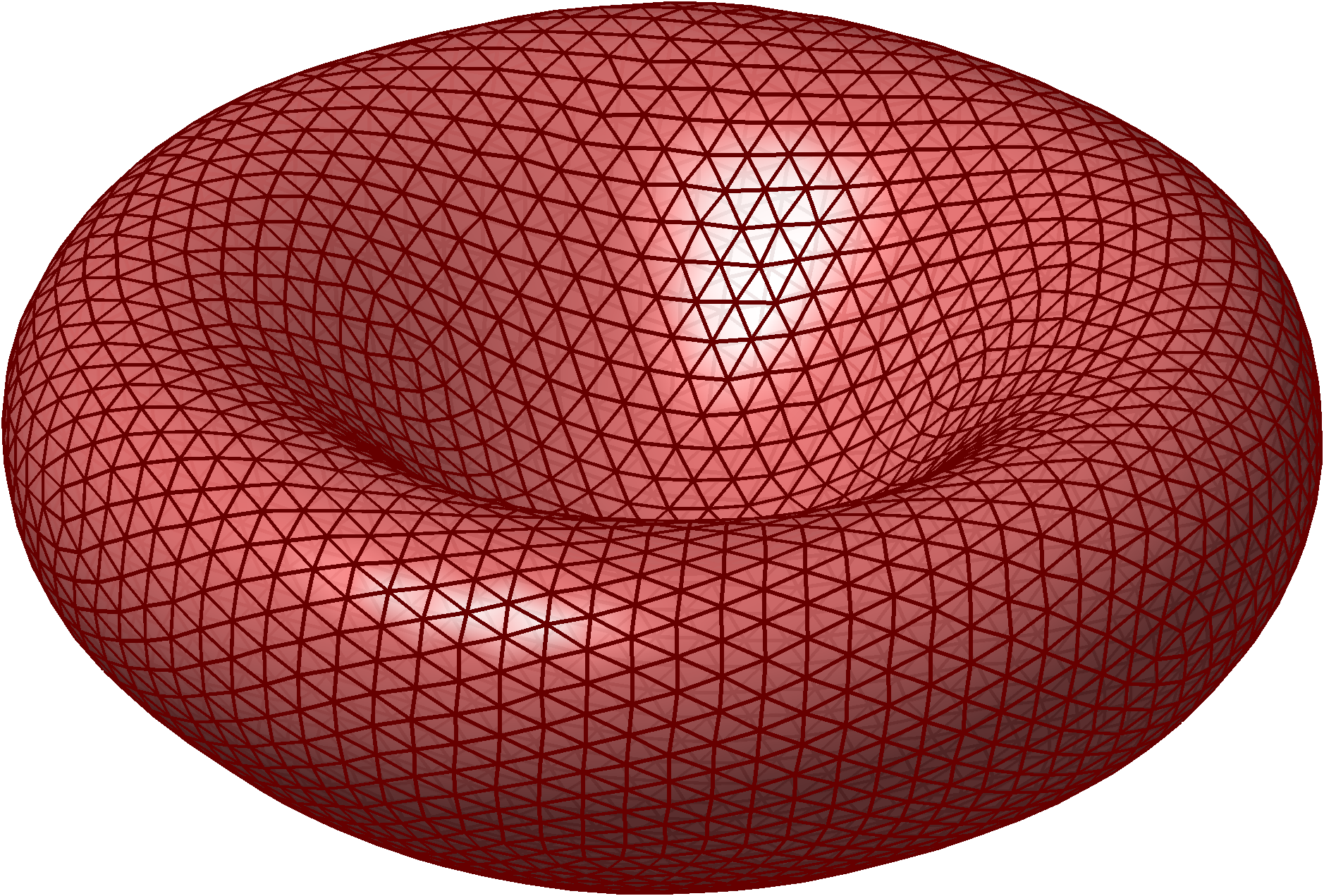}}  \\[11pt]
Sto.I.2 & 96.12 & 200.0 & 16 & 132.57 & 91.75 & 0.0 & 4,900 & 5,000 & 5,000 & \adjustbox{valign=m}{\includegraphics[width=0.095\textwidth]{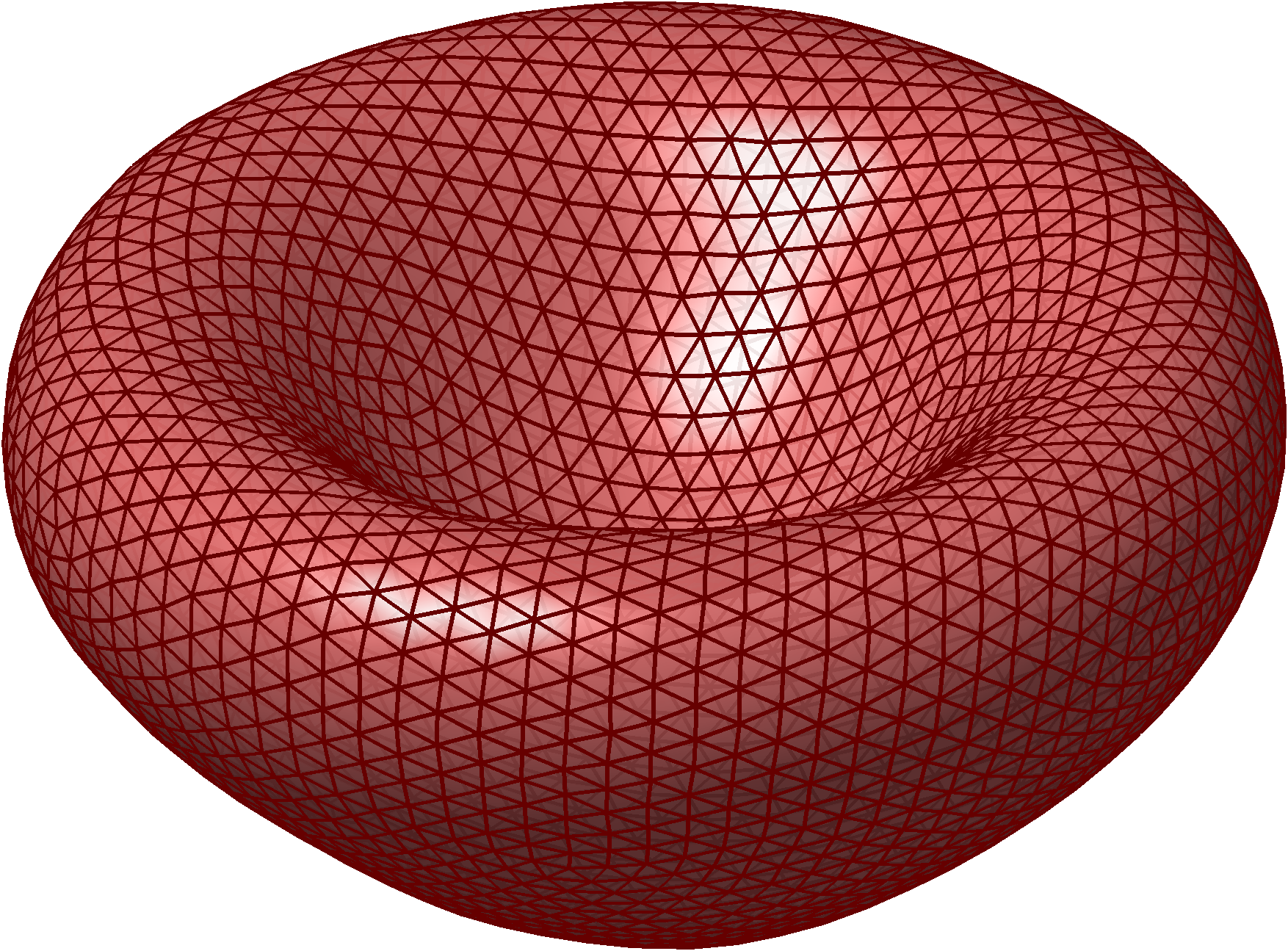}}  \\[11pt]
Sto.II.1 (Repre. Sto.II) & 96.12 & 400.0 & 16 & 132.57 & 91.75 & 0.0 & 4,900 & 5,000 & 5,000 & \adjustbox{valign=m}{\includegraphics[width=0.09\textwidth]{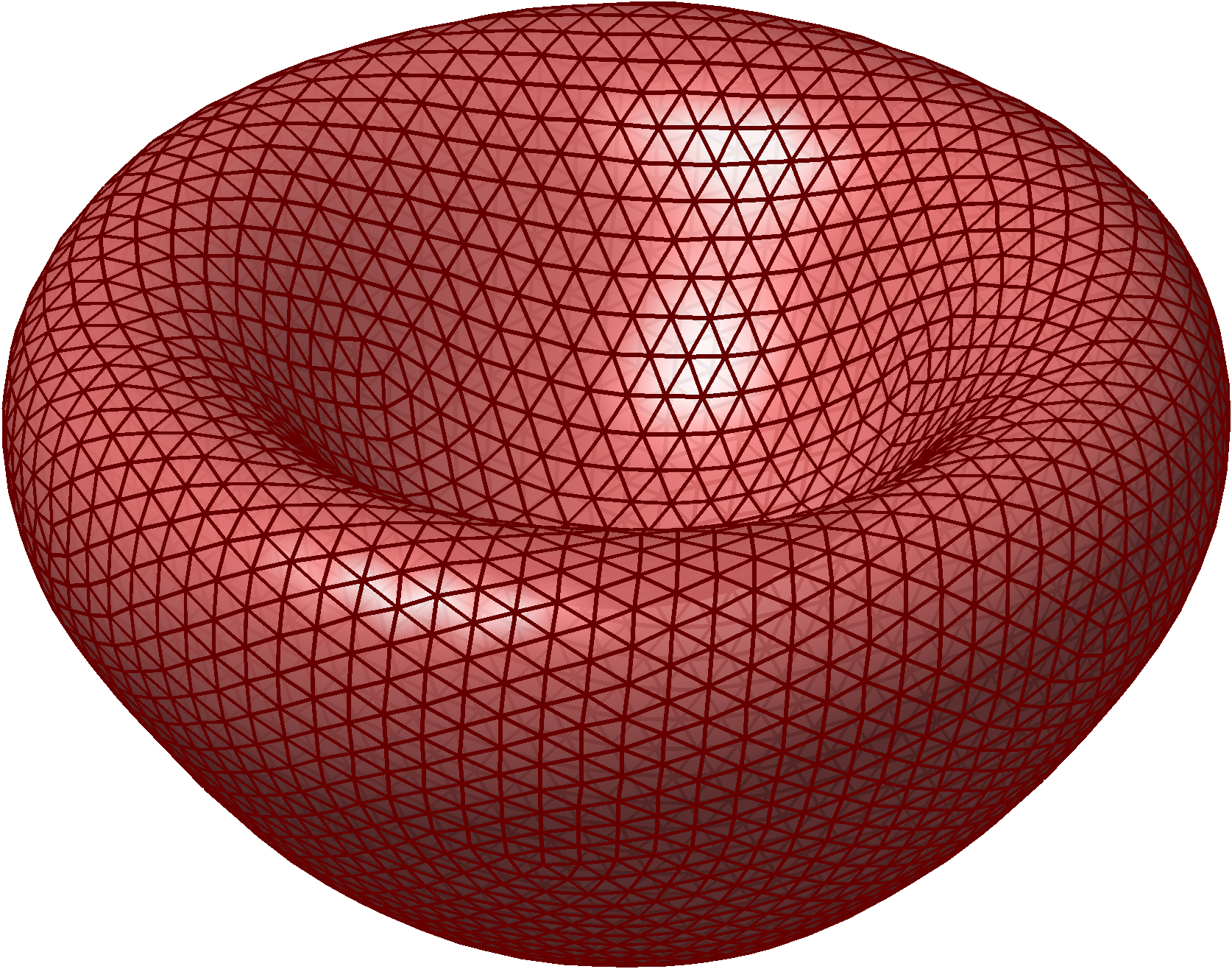}}  \\[11pt]
Sto.II.2 & 96.12 & 800.0 & 16 & 132.57 & 91.75 & 0.0 & 4,900 & 5,000 & 5,000 & \adjustbox{valign=m}{\includegraphics[width=0.085\textwidth]{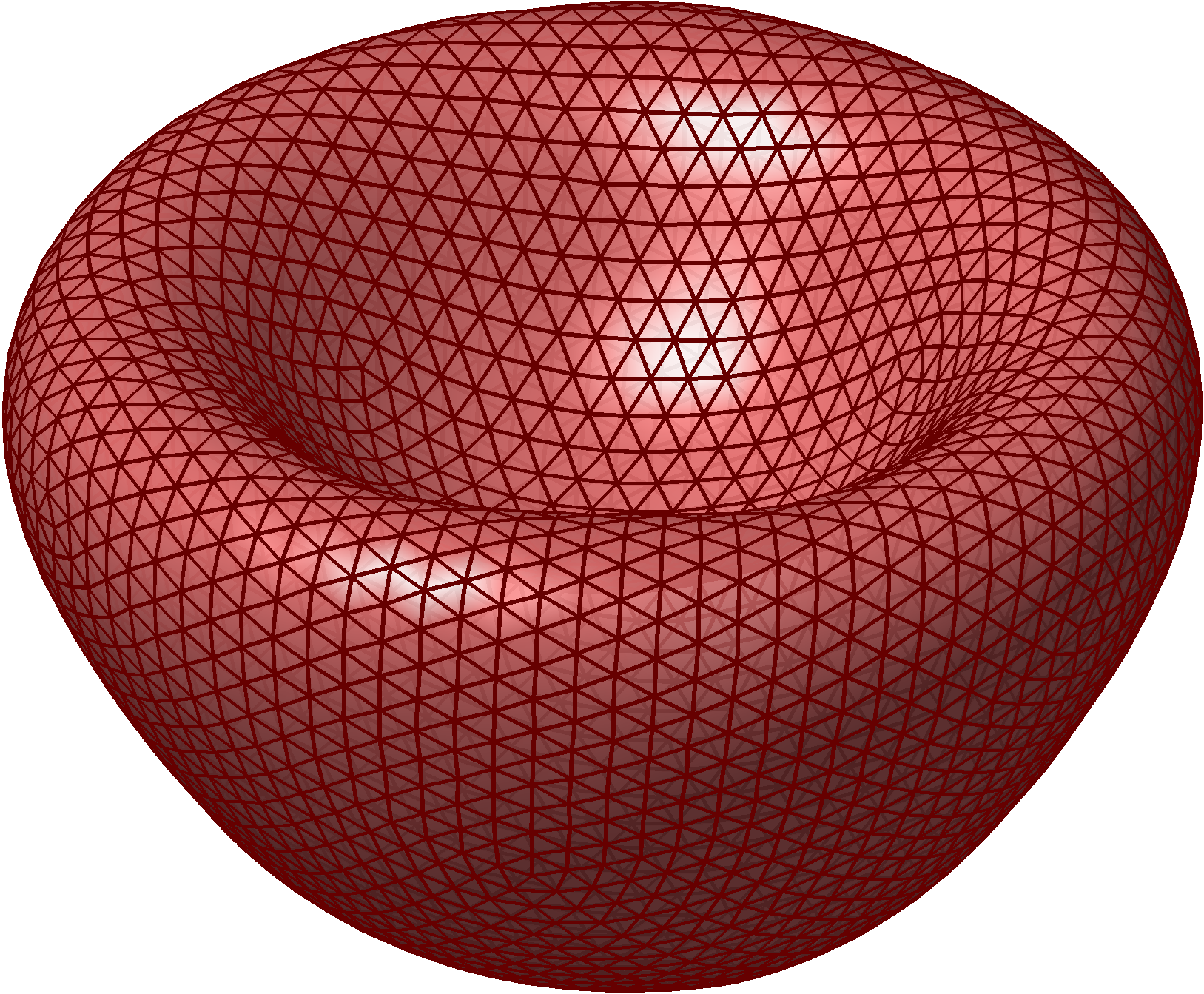}}  \\[11pt]
Ech.I.1 (Repre. Ech.I) & 96.12 & 6.44 & 1 & 132.57 & 92.45 & 0.113 & 10,000 & 10,000 & 5,000 & \adjustbox{valign=m}{\includegraphics[width=0.1\textwidth]{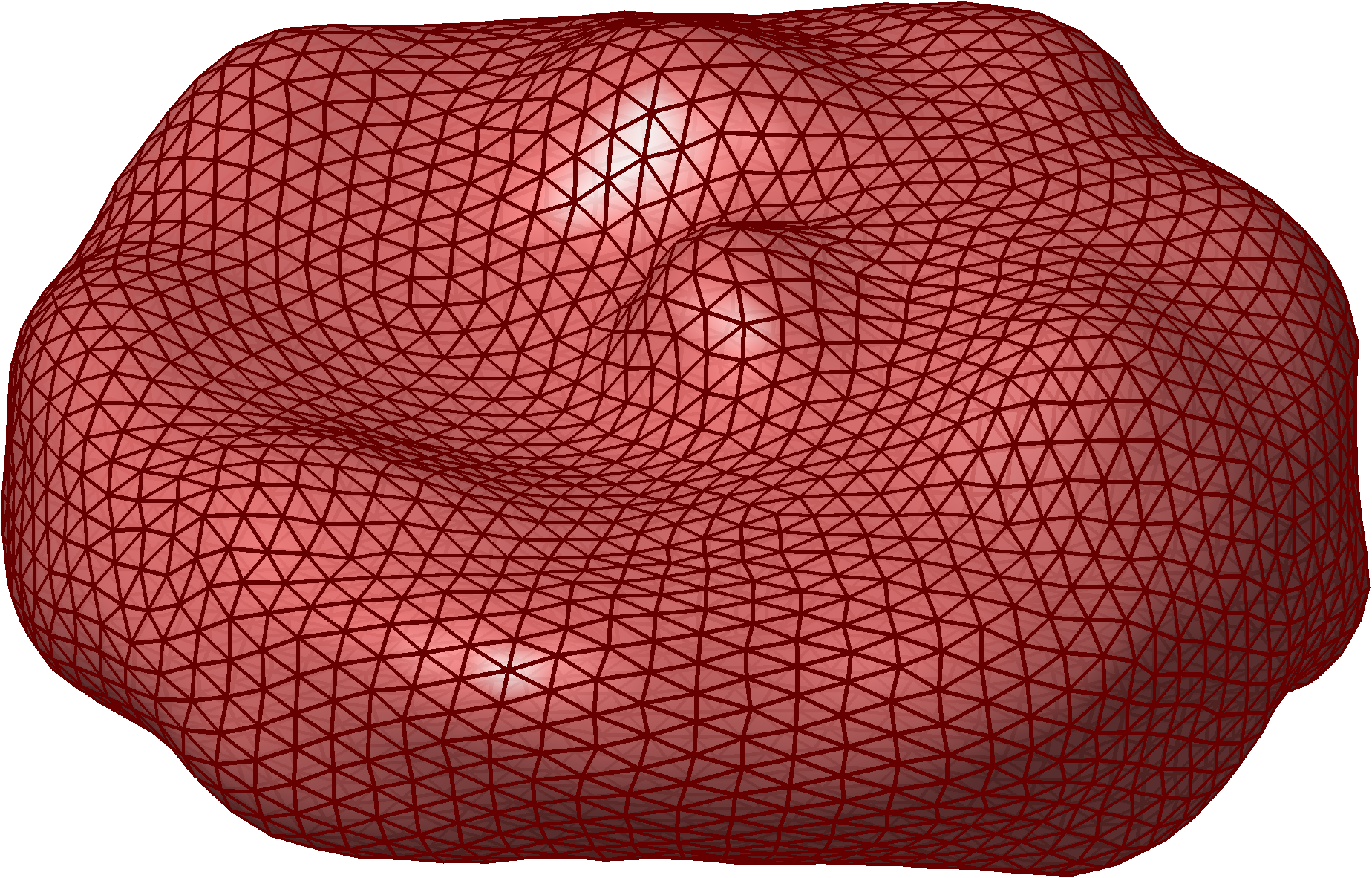}} \\[10pt]
Ech.I.2 & 96.12 & 20.0 & 1 & 132.57 & 92.45 & 0.113 & 10,000 & 10,000 & 5,000 & \adjustbox{valign=m}{\includegraphics[width=0.1\textwidth]{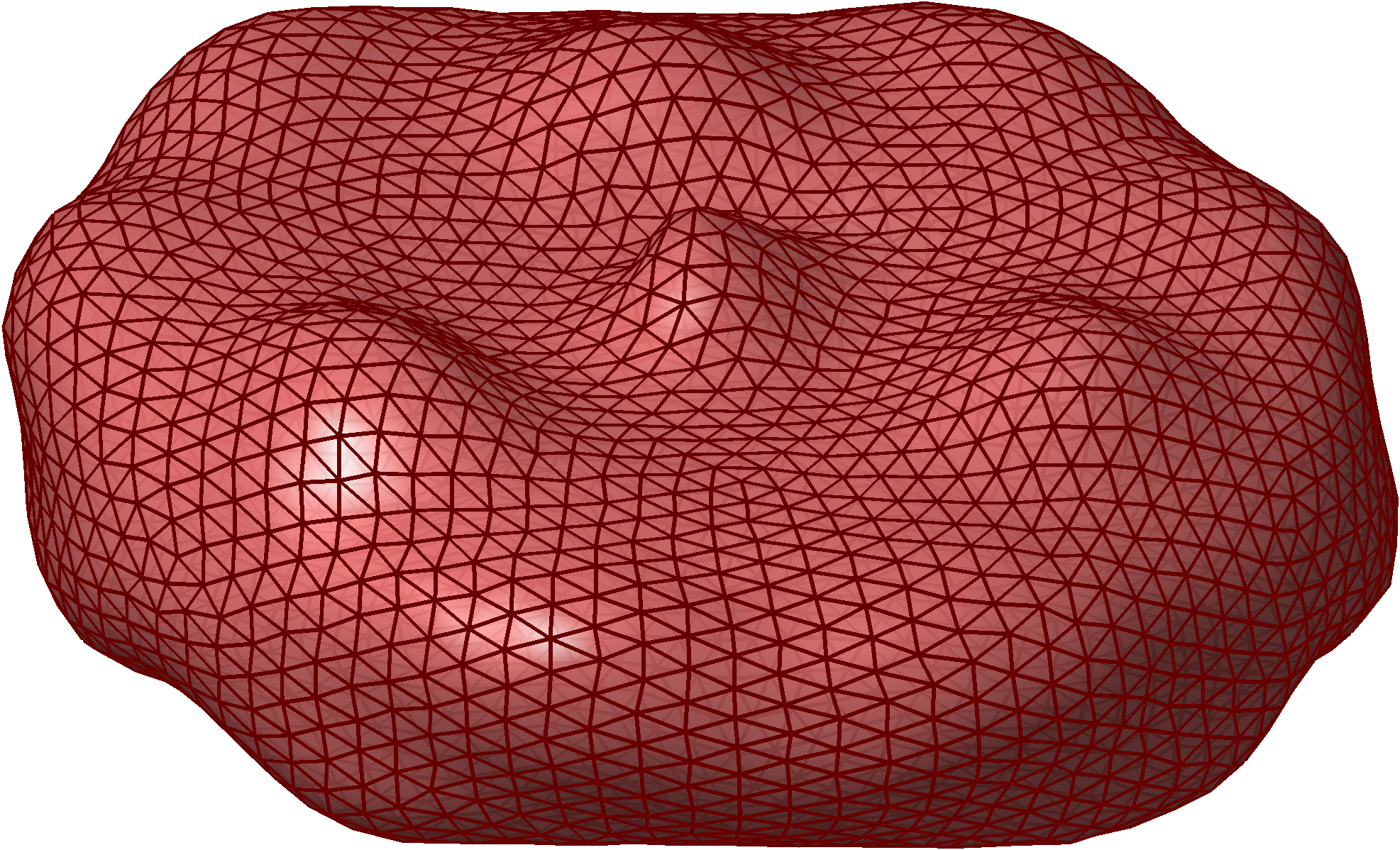}} \\[10pt]
Ech.I.3 & 96.12 & 20.0 & 1 & 125.95 & 92.45 & 0.066 & 10,000 & 10,000 & 5,000 & \adjustbox{valign=m}{\includegraphics[width=0.095\textwidth]{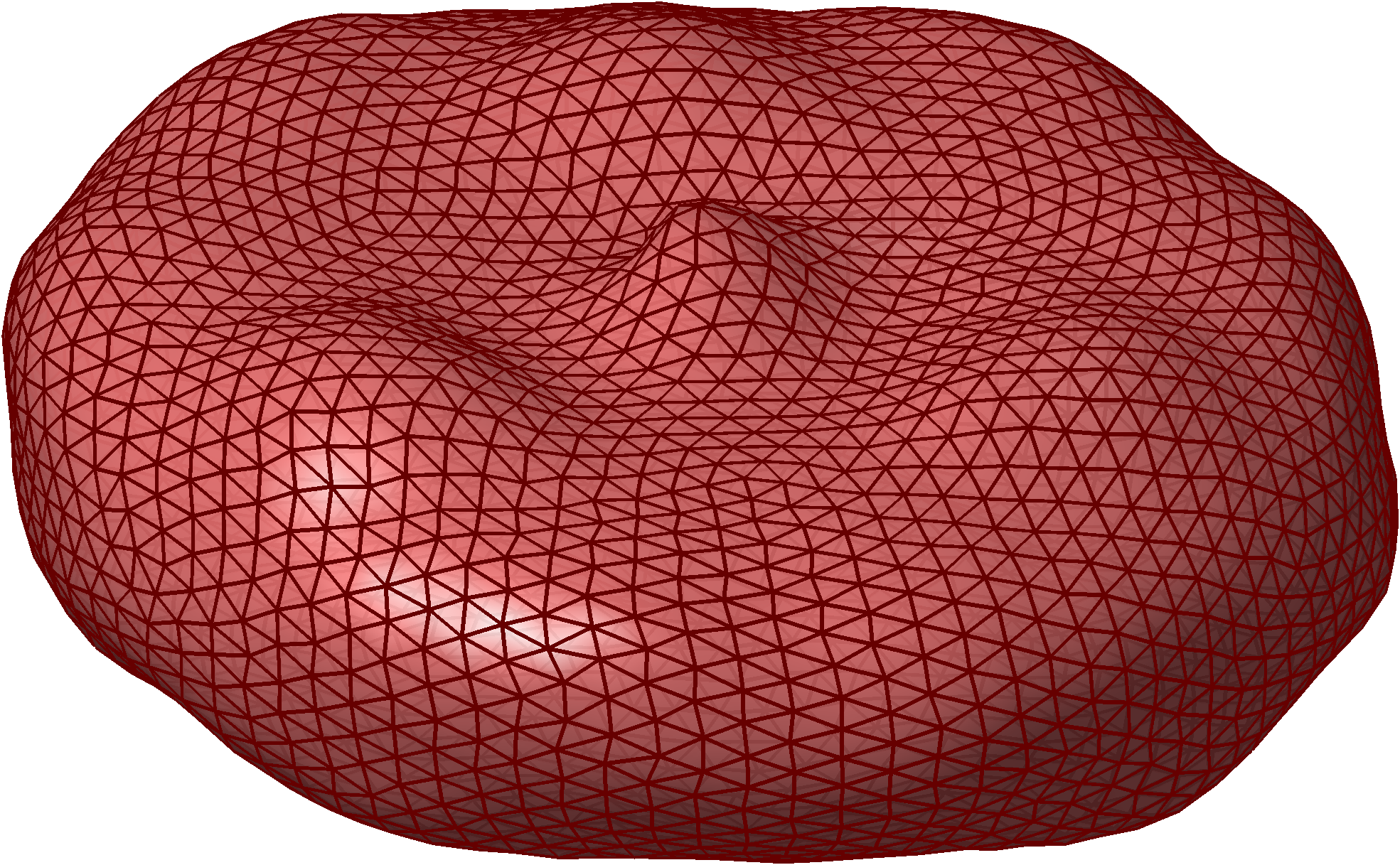}} \\[10pt]
Ech.II.1 (Repre. Ech.II) & 96.12 & 50.0 & 1 & 132.57 & 92.45 & 0.226 & 10,000 & 10,000 & 5,000 & \adjustbox{valign=m}{\includegraphics[width=0.08\textwidth]{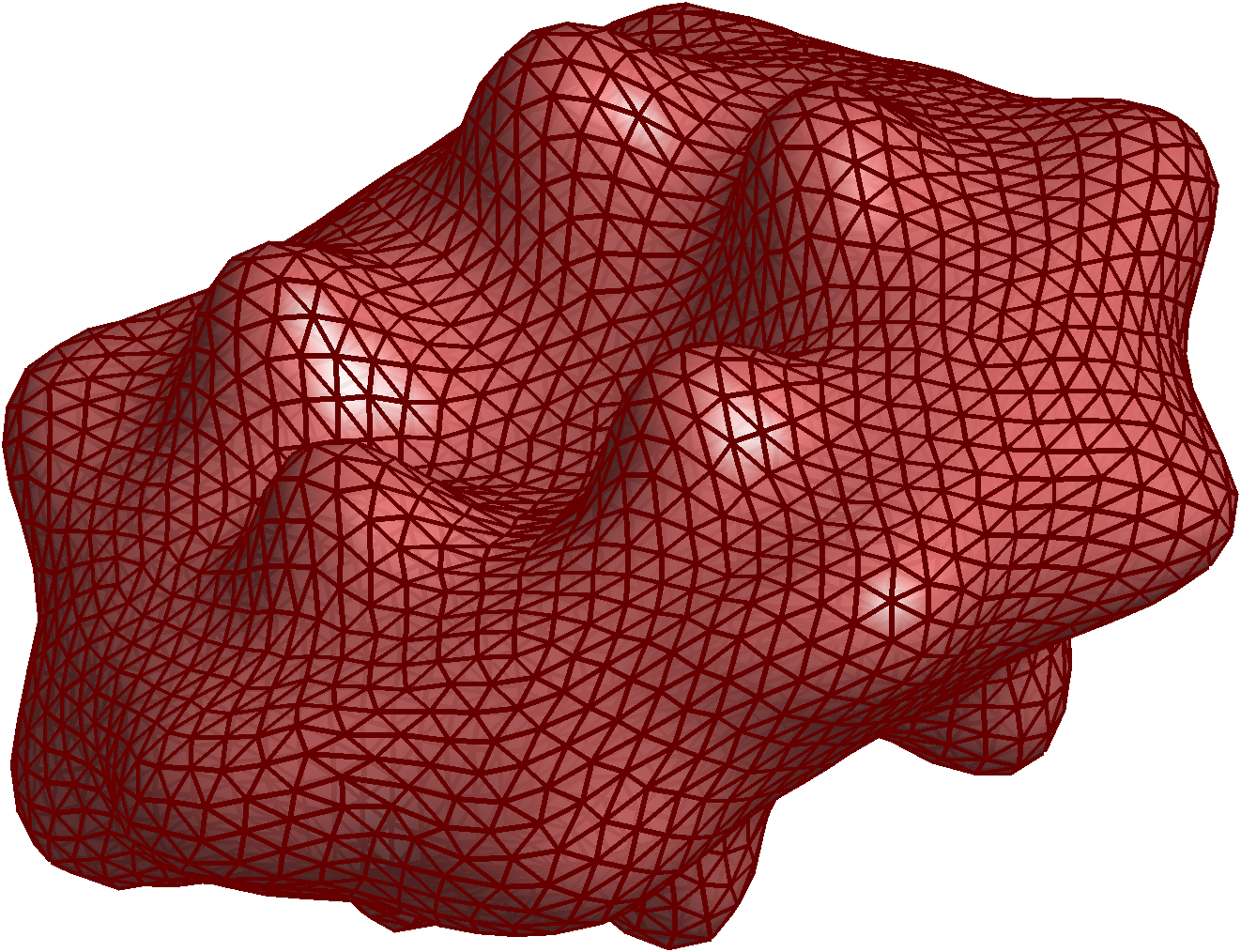}} \\[10pt]
Ech.II.2 & 96.12 & 90.0 & 1 & 132.57 & 92.45 & 0.226 & 10,000 & 10,000 & 5,000 & \adjustbox{valign=m}{\includegraphics[width=0.08\textwidth]{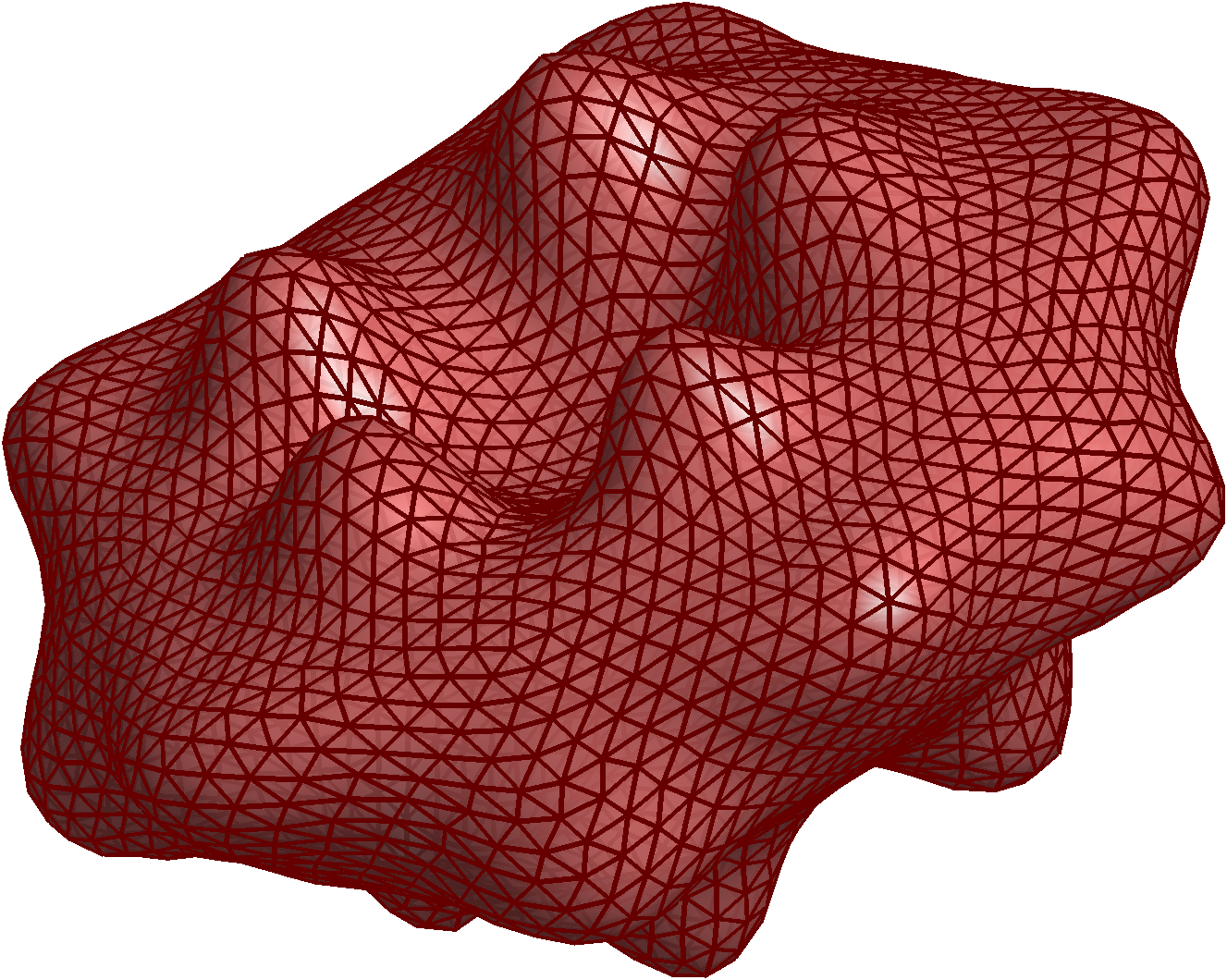}} \\[10pt]
Ech.II.3 & 96.12 & 15.0 & 1 & 125.95 & 92.45 & 0.226 & 10,000 & 10,000 & 5,000 & \adjustbox{valign=m}{\includegraphics[width=0.076\textwidth]{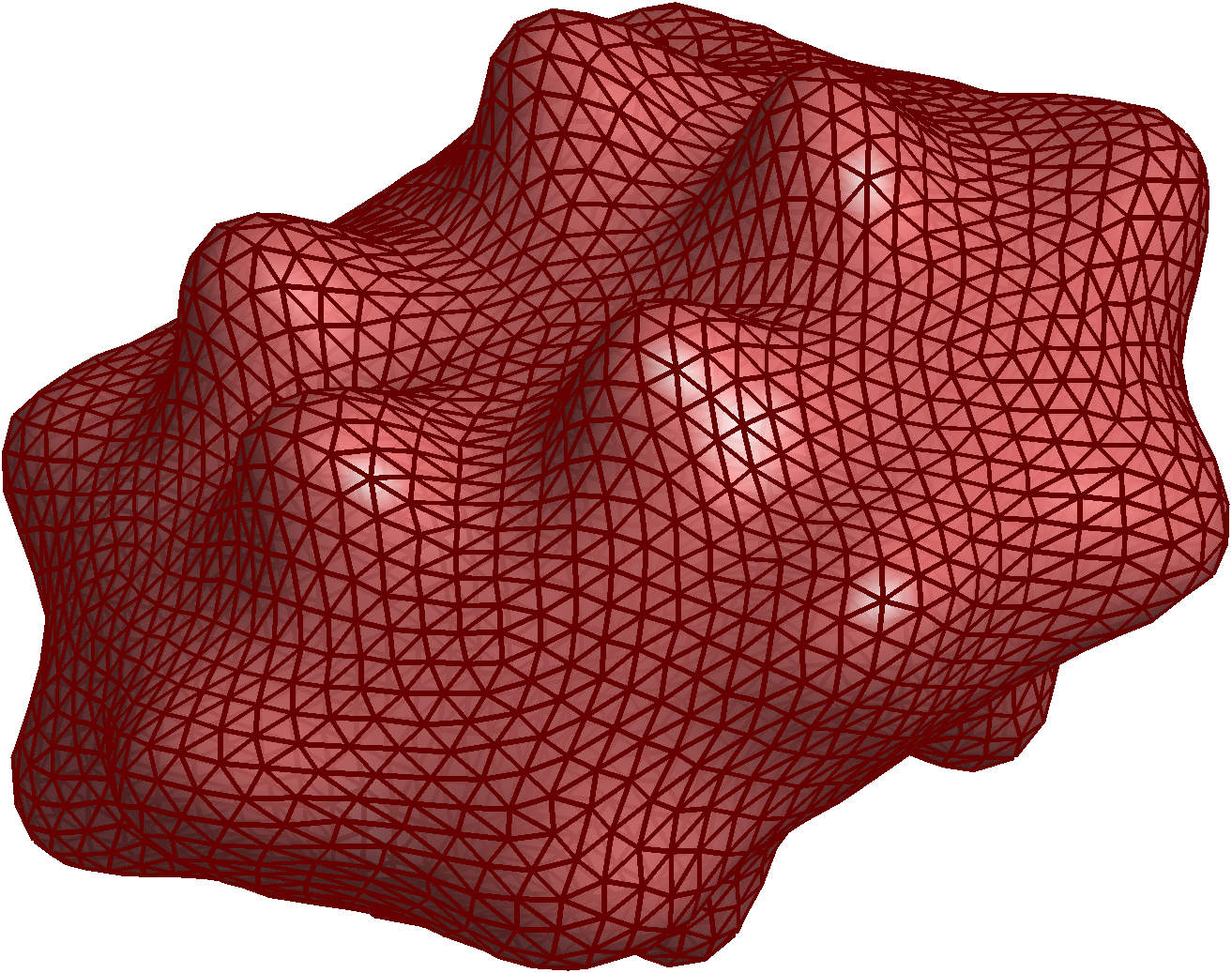}} \\[10pt]
Ech.III.1 (Repre. Ech.III) & 96.12 & 150.0 & 1 & 132.57 & 92.45 & 0.377 & 10,000 & 10,000 & 5,000 & \adjustbox{valign=m}{\includegraphics[width=0.09\textwidth]{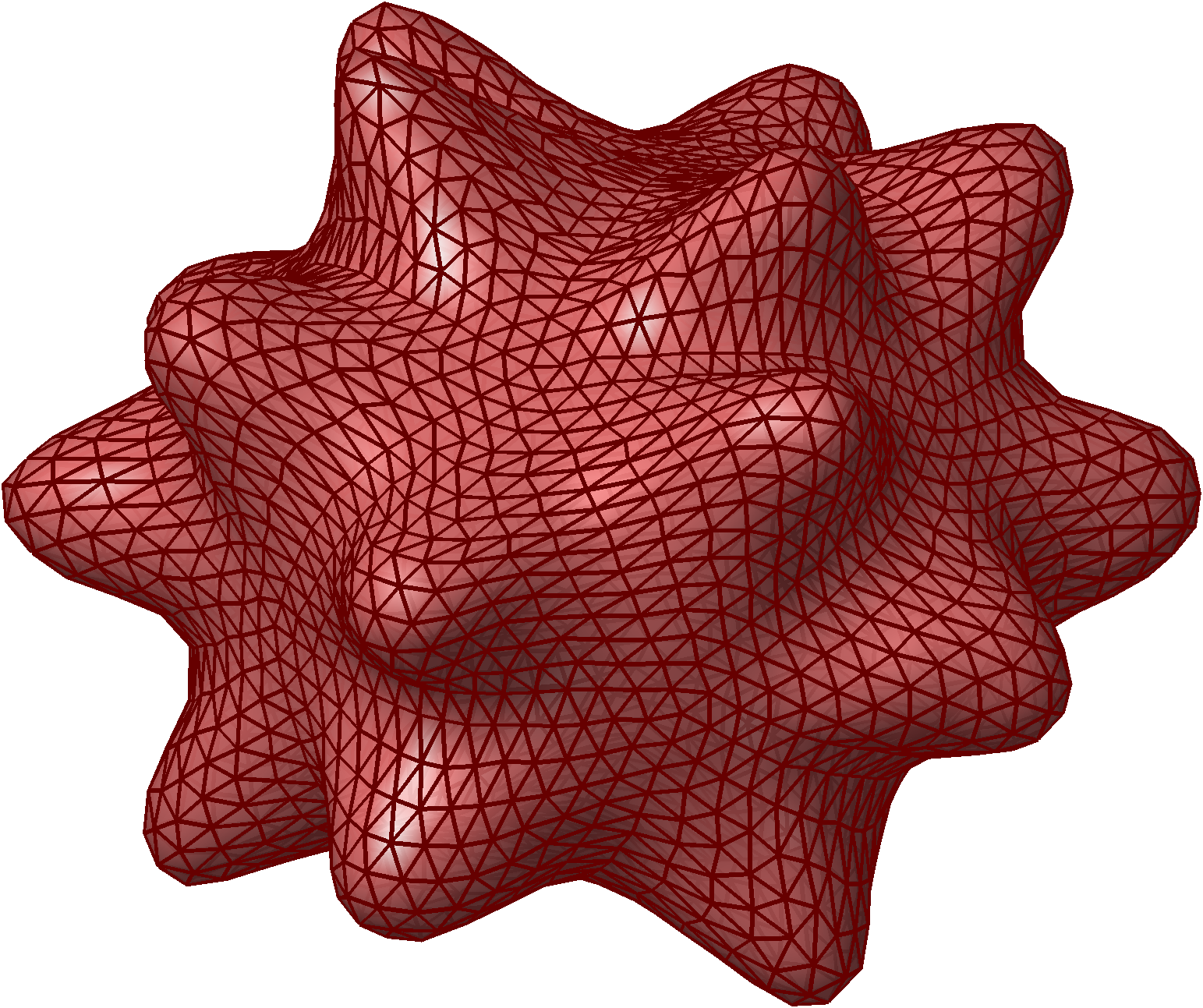}} \\[10pt]
Ech.III.2 & 96.12 & 200.0 & 1 & 125.95 & 92.45 & 0.344 & 10,000 & 10,000 & 5,000 & \adjustbox{valign=m}{\includegraphics[width=0.0855\textwidth]{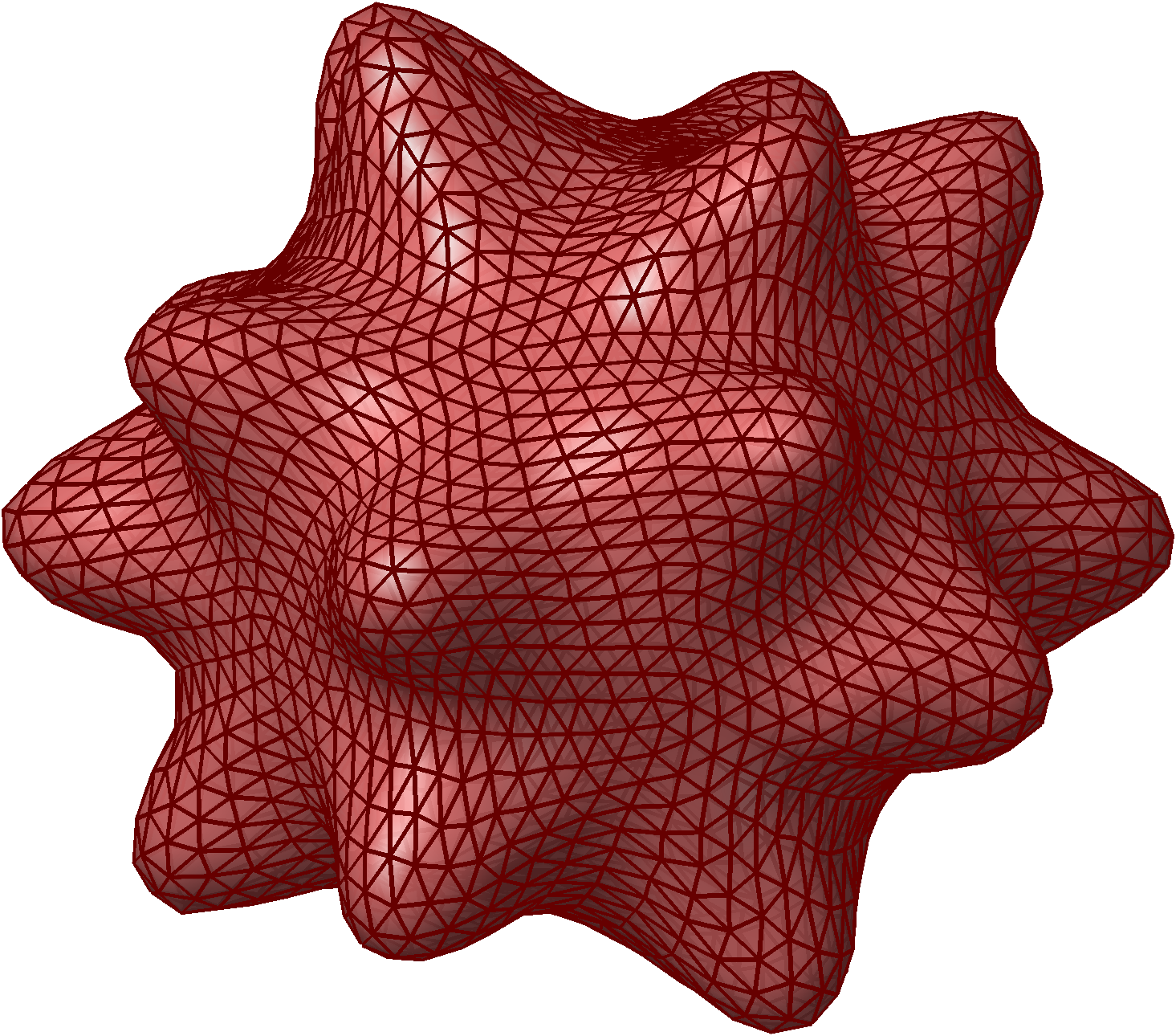}} \\[10pt]
Ech.III.3 & 96.12 & 200.0 & 1 & 119.32 & 92.45 & 0.308 & 10,000 & 10,000 & 5,000 & \adjustbox{valign=m}{\includegraphics[width=0.081\textwidth]{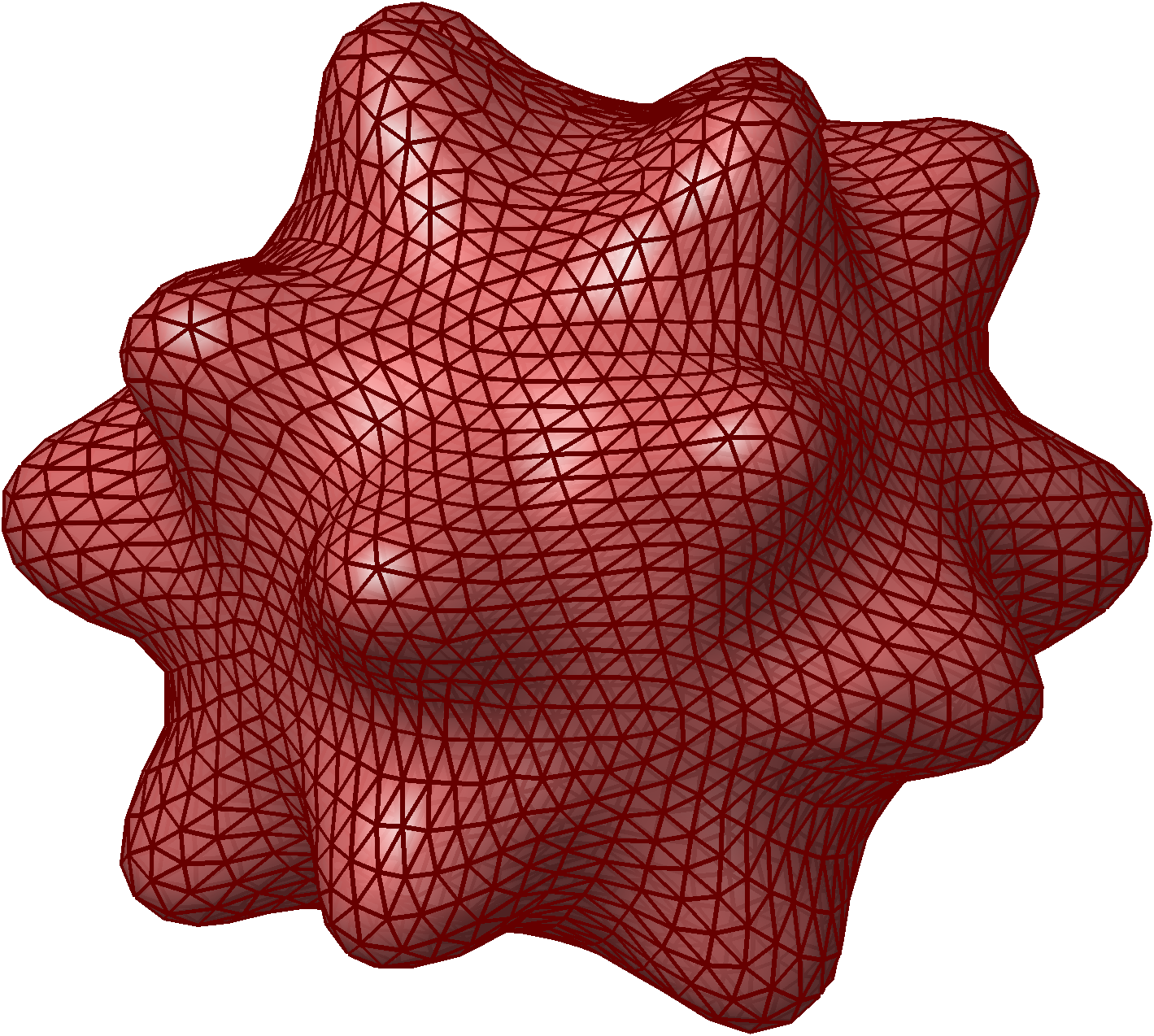}} \\[10pt]
\bottomrule
\end{tabularx}
\end{table}

\section{Optimization of the rotation matrix}\label{app:B}
Free rotation of RBCs is commonly observed in various settings, such as in vivo microcirculation~\cite{Kaul01012004}, in vitro flow chambers~\cite{Sugii_2005}, and optical trapping experiments~\cite{zhong2013trapping}, where cells are suspended in fluid and unconstrained by fixed orientation. Thus, the free rotation is a vital factor that needs to be considered during the imaging. Since our surrogate model conducts prediction using the coordinate-based datasets, the misalignment in the identical rotational angle between the target shape and the base shape will significantly affect the prediction result. Therefore, we developed a simple optimization procedure to ensure the rotational alignment before the training of the MFNN. 

Given the base shape in the coordinate cluster $X^\text{base}$ and cross-sectional information of the target shape in the coordinate cluster $X^\text{target}$, the discrepancy between the two clusters can be calculated using the chamfer distance~\cite{5539837}, which is commonly utilized to quantify the similarity among point clusters. The root chamfer distance $d_\text{chamfer}$ is mathematically expressed as
\begin{equation}
 d_\text{chamfer} = D_\text{cf}(X^{\text{base}},X^{\text{target}}) = \sqrt{ \frac{\text{1}}{|X^\text{base}|} \sum_{\mathbf{x^\text{b}} \in X^\text{base}} \min_{\mathbf{x^\text{t}} \in X^\text{target}} \| \mathbf{x^\text{b}} - \mathbf{x^\text{t}} \|_\text{2} + \frac{1}{|X^\text{target}|} \sum_{\mathbf{x^\text{t}} \in X^\text{target}} \min_{\mathbf{x^\text{b}} \in X^\text{base}} \| \mathbf{x^\text{b}} - \mathbf{x^\text{t}} \|_\text{2}},   
\end{equation}
where $X^{\text{base}} = \{ \mathbf{x}^{\text{b}}_{\text{1}}, \mathbf{x}^{\text{b}}_{\text{2}}, \dots, \mathbf{x}^{\text{b}}_{\text{n}} \}$ and $X^{\text{target}} = \{ \mathbf{x}^{\text{t}}_{\text{1}}, \mathbf{x}^{\text{t}}_{\text{2}}, \dots, \mathbf{x}^{\text{t}}_{\text{m}}\}$.

Assuming a rotation matrix $R$ that performs an undetermined rotation operation to $X^{\text{target}}$, the rotated cluster $X^{\text{rot-target}}$ is
\begin{equation}
    X^{\text{rot-rotated}} =R\otimes X^{\text{target}}= \{ R \mathbf{x}^{\text{t}}_{\text{1}}, R \mathbf{x}^{\text{t}}_{\text{2}}, \dots, R \mathbf{x}^{\text{t}}_{\text{m}}\}
\end{equation}

The optimal rotation matrix \( R^* \) is then obtained by solving the following optimization problem:
\begin{equation}
    R^* = \arg\min_{R \in SO(3)} D_{\text{cf}}(X^{\text{base}}, R \otimes X^{\text{target}}),
    \label{eq:rotation_opt}
\end{equation}
where \( SO(3) \) denotes the special orthogonal group of 3D rotation matrices. In practice, we parameterize the rotation matrix \( R \) using a rotation vector (axis–angle representation) and solve the optimization problem using a specific gradient-based method using Broyden–Fletcher–Goldfarb–Shanno (BFGS) algorithm~\cite{liu1989limited}.

Although this simple optimization procedure effectively ensures rotational alignment across all training and testing RBC samples in our current study, it is important to recognize that real-world RBC orientations are often more complex. Therefore, a more robust and generalizable approach should be developed to handle rotational alignment in practical scenarios.

\section{Setups for low-fidelity CNN validation and MFNN predictor}\label{app:C}
\subsection{Neural network setups}

\begin{table}[htbp]
\centering
    \caption{CNN setups of six representative cases in validation benchmark}
\label{tab:CNN}
\begin{tabularx}{\textwidth}{c|c c c c}
\toprule
\textbf{Shape} & NN structure & Batch & Total epochs & Training strategy\\
\midrule
Dis. & \makecell[l]{CNN component:\\{[3] + [32] + [64] + [128]};\\ FNN component:\\{[128] + [50]*3 + [3]}.} & 128 & 400,000 & \makecell[l]{1. Adam with $lr=\text{10}^\text{-4}$ for 200,000 epochs,\\2. SGD with Nesterov momentum $\gamma=0.9$,\\ \quad $lr=\text{10}^\text{-5}$ for 200,000 epochs.}\\ \hline
\makecell[c]{\rule{0pt}{3ex}Sto.I,\\ Sto.II,\\ Ech.I,\\ Ech.II,\\ Ech.III} & \makecell[l]{CNN componet:\\{[3] + [32] + [64] + [128] + [256]};\\ FNN component:\\{[256] + [100]*3 + [3]}.} & 128 & 600,000 & \makecell[l]{1. Adam with $lr=\text{10}^\text{-4}$ for 300,000 epochs,\\2. SGD with Nesterov momentum $\gamma=0.9$,\\ \quad $lr=\text{10}^\text{-5}$ for 300,000 epochs.}\\
\bottomrule
\end{tabularx}
\end{table}

Six representative cases have been investigated to validate the effectiveness of low-fidelity CNN, and detailed setups are shown in Table.~\ref{tab:CNN}. Regarding the result cases in the MFNN predictor, all the detailed setups are shown in Table.~\ref{tab:MFNN}. The training process of all cases is performed on a NVIDIA RTX4090 GPU platform with Tensorflow-2.18.0, and the average training time of a low-fidelity CNN validation case and an MFNN prediction case is 21.89 minutes and 35.65 minutes, respectively. 

\begin{table}[htbp]
\centering
\caption{MFNN setups of all the cases in the prediction}
\label{tab:MFNN}
\begin{tabularx}{\textwidth}{c|c|c|>{\centering\arraybackslash}X}
\toprule
\textbf{Case id} & Base shape & Target shape & Shared training setup \\
\midrule
\textit{D.D.(1)} & Dis.1 & Dis.2 & 
\multirow{13}{\linewidth}{
%\centering
\begin{minipage}[c]{\linewidth}
\centering
Low-fidelity CNN component: [3] + [32] + [64] + [128] + [256];\\ 
Low-fidelity FNN component: [256] + [100]*3 + [3];\\
High-fidelity nonlinear FNN component: [6] + [100]*3 + [3];\\
High-fidelity linear FNN component: [6] + [3].\\ \  \\
Weights in Loss function: $w_\text{1}=0.1$, $w_\text{2}=0.01$, and $\lambda=0.1$.\\
Batch for Low-fidelity CNN component: 128\\
Total epochs: 400,000\\ \ \\
Training strategy:\\ 
1. Adam with $lr=\text{10}^\text{-4}$ for 200,000 epochs;\\2. SGD with Nesterov momentum $\gamma=0.9$, $lr=\text{10}^\text{-5}$ for 200,000 epochs.
\end{minipage}
} \\
\cline{1-3}
\textit{D.D.(2)} & Dis.1 & Dis.3 & \\
\cline{1-3}
\textit{S.S.(1)} & Sto.I.1 & Sto.I.2 & \\
\cline{1-3}
\textit{S.S.(2)} & Sto.I.1 & Sto.II.1 & \\
\cline{1-3}
\textit{S.S.(3)} & Sto.I.1 & Sto.II.2 & \\
\cline{1-3}
\textit{D.S.} & Dis.1 & Sto.I.1 & \\
\cline{1-3}
\textit{E.E.(1)} & Ech.I.1 & Ech.I.2 & \\
\cline{1-3}
\textit{E.E.(2)} & Ech.I.1 & Ech.I.3 & \\
\cline{1-3}
\textit{E.E.(3)} & Ech.II.1 & Ech.II.2 & \\
\cline{1-3}
\textit{E.E.(4)} & Ech.II.1 & Ech.II.3 & \\
\cline{1-3}
\textit{E.E.(5)} & Ech.III.1 & Ech.III.2 & \\
\cline{1-3}
\textit{E.E.(6)} & Ech.III.1 & Ech.III.3 & \\
\cline{1-3}
\textit{E.E.(7)} & Ech.I.1 & Ech.II.1 & \\
\cline{1-3}
\textit{E.E.(8)} & Ech.II.2 & Ech.III.3 & \\
\bottomrule
\end{tabularx}
\end{table}

\subsection{Sampling setups}
The cross-sectional sample dataset for training is formed by vertices located in three orthogonal cross-sections with small thickness $\delta$ ($\delta<<R_\text{c}$, where $R_\text{c}$ is the characteristic radius of RBC) in Cartesian space. The small size and cross-sectional dependence of the training set are set to mimic the limited observation at several fixed angles. The demonstration of typical cross-sectional sampling is shown in Fig.~\ref{fig:M3-crossdemo}. The number of sample points obtained by cross-sectional sampling of all RBC cases is shown in Table.~\ref{tab:cs-sampling}. We have also investigated the effect of I.O.C. sampling on prediction performance in the echinocyte scenario. The number of sample points obtained by I.O.C. sampling of the echinocyte cases is shown in Table.~\ref{tab:ioc-sampling}.

\begin{table}[htbp]
\centering
    \caption{The number of sample points obtained by cross-sectional sampling of all RBC cases, $\delta$ is uniformly set as 0.1$\mu\text{m}$}
\label{tab:cs-sampling}
\begin{tabularx}{\textwidth}{c|c c c c c c c}
\toprule
\textbf{Shape id} & 3.C.S.s & 2.C.S.s:yz+xz & 2.C.S.s:xz+xy & 2.C.S.s:yz+xy & 1.C.S.:yz & 1.C.S.:xz & 1.C.S.:xy \\
\midrule
Dis.1 & 231 & 136 & 165 & 165 & 69 & 69 & 97 \\
Dis.2 & 228 & 136 & 160 & 165 & 71 & 67 & 95 \\
Dis.3 & 238 & 143 & 170 & 168 & 71 & 74 & 98 \\
Sto.I.1 & 242 & 148 & 168 & 171 & 75 & 73 & 97 \\
Sto.I.2 & 244 & 153 & 167 & 173 & 80 & 75 & 94 \\
Sto.II.1 & 245 & 157 & 169 & 169 & 79 & 80 & 91 \\
Sto.II.2 & 246 & 159 & 168 & 170 & 81 & 80 & 90 \\
Ech.I.1 & 251 & 155 & 178 & 175 & 77 & 80 & 100 \\
Ech.I.2 & 242 & 149 & 167 & 173 & 78 & 73 & 96 \\
Ech.I.3 & 244 & 151 & 174 & 170 & 74 & 79 & 98 \\
Ech.II.1 & 203 & 169 & 110 & 131 & 97 & 74 & 36 \\
Ech.II.2 & 219 & 179 & 121 & 142 & 102 & 79 & 42 \\
Ech.II.3 & 229 & 191 & 131 & 140 & 102 & 91 & 40 \\
Ech.III.1 & 263 & 174 & 174 & 184 & 93 & 83 & 93 \\
Ech.III.2 & 271 & 184 & 176 & 188 & 99 & 87 & 91 \\
Ech.III.3 & 259 & 174 & 174 & 175 & 88 & 88 & 88\\
\bottomrule
\end{tabularx}
\end{table}

\begin{table}[htbp]
\centering
    \caption{Demonstration and the sample points number of I.O.C sampling in \textit{E.E.(8)} case}
\label{tab:ioc-sampling}
\begin{tabularx}{\textwidth}{c|c c c c c c}
\toprule
\  & Pattern 1 & Pattern 2 & Pattern 3 & Pattern 4 & Pattern 5 & Pattern 6 \\
\midrule
Demonstration & \adjustbox{valign=m}{\includegraphics[width=0.1\textwidth]{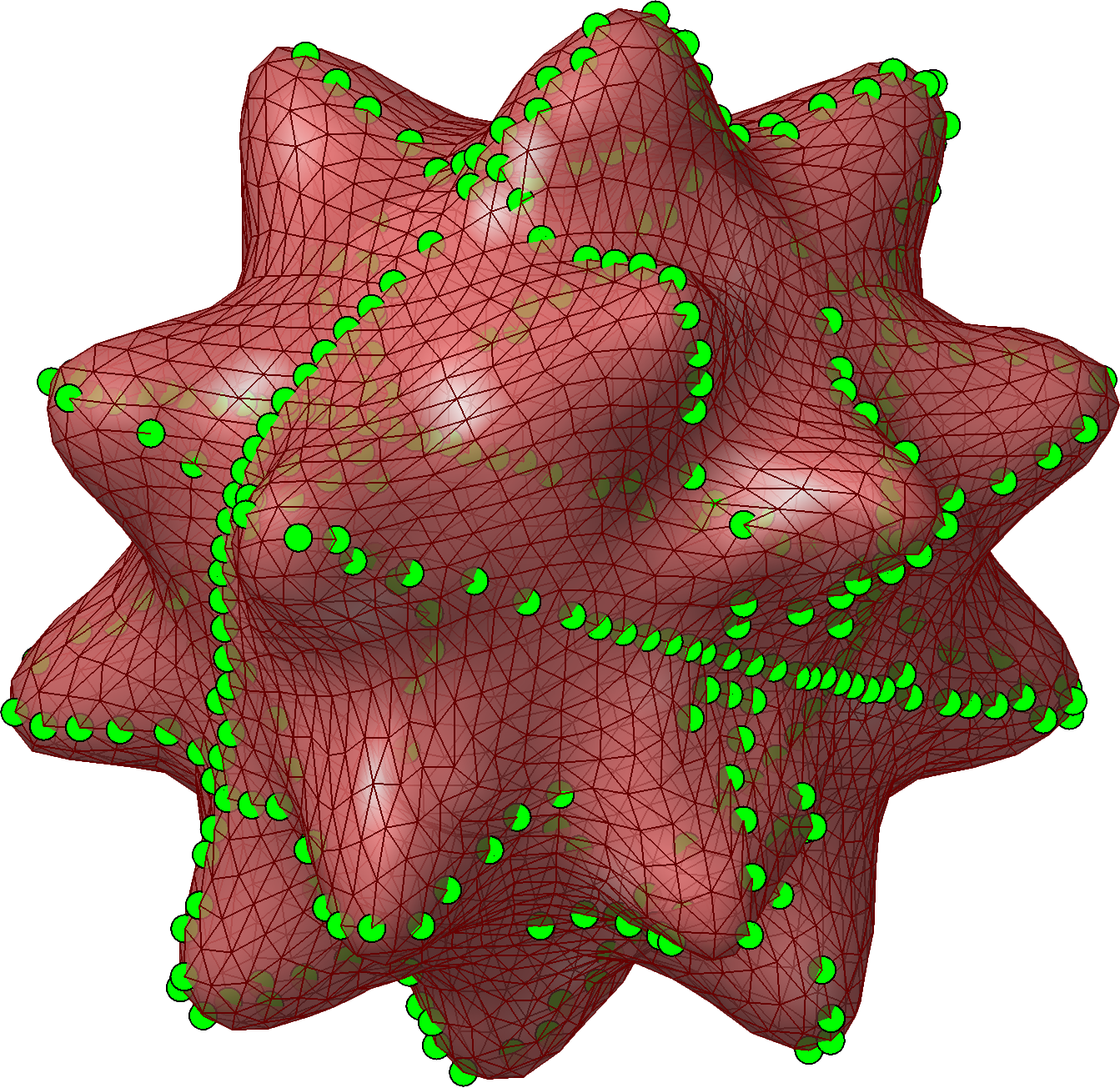}} & \adjustbox{valign=m}{\includegraphics[width=0.1\textwidth]{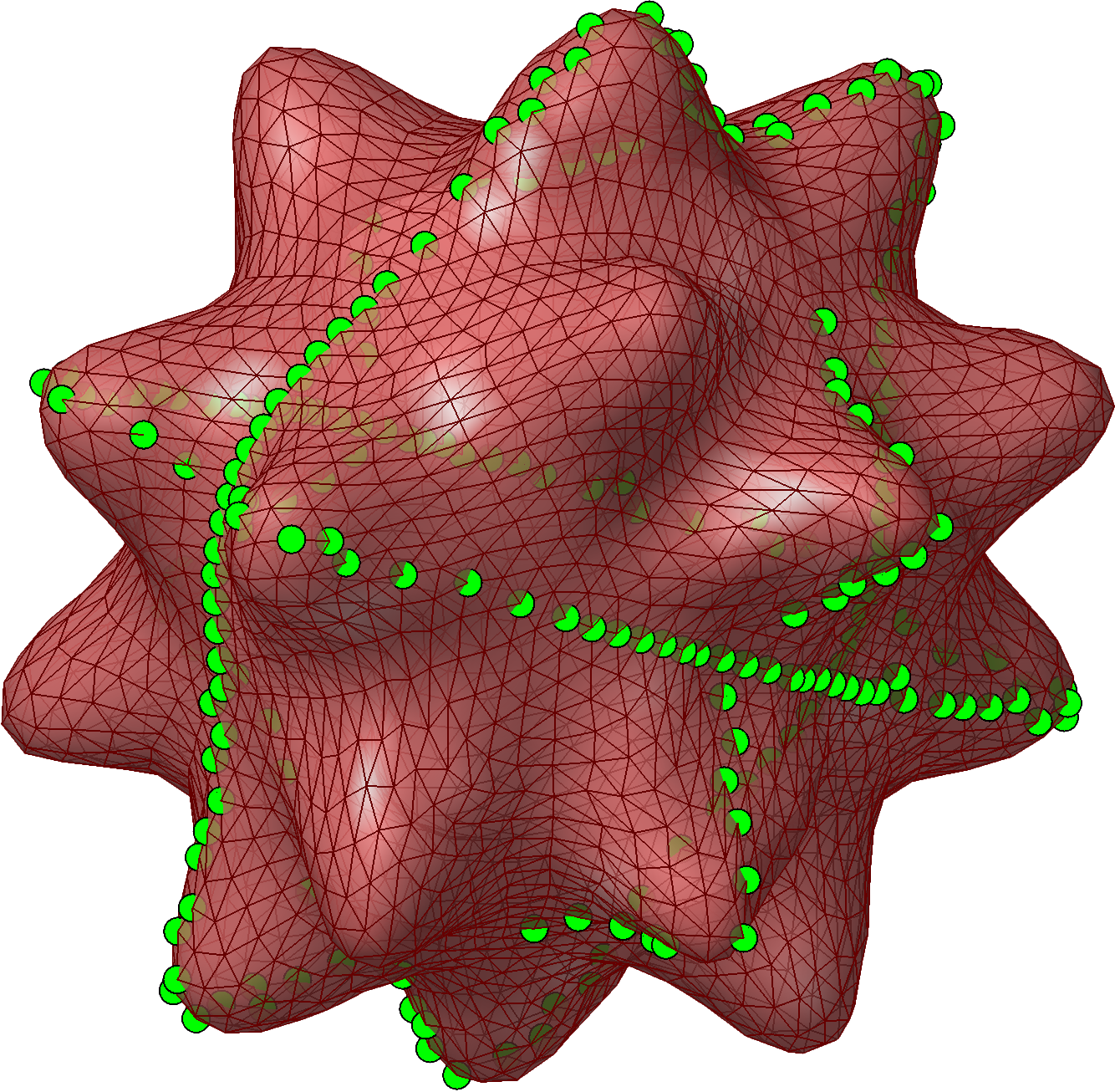}} & \adjustbox{valign=m}{\includegraphics[width=0.1\textwidth]{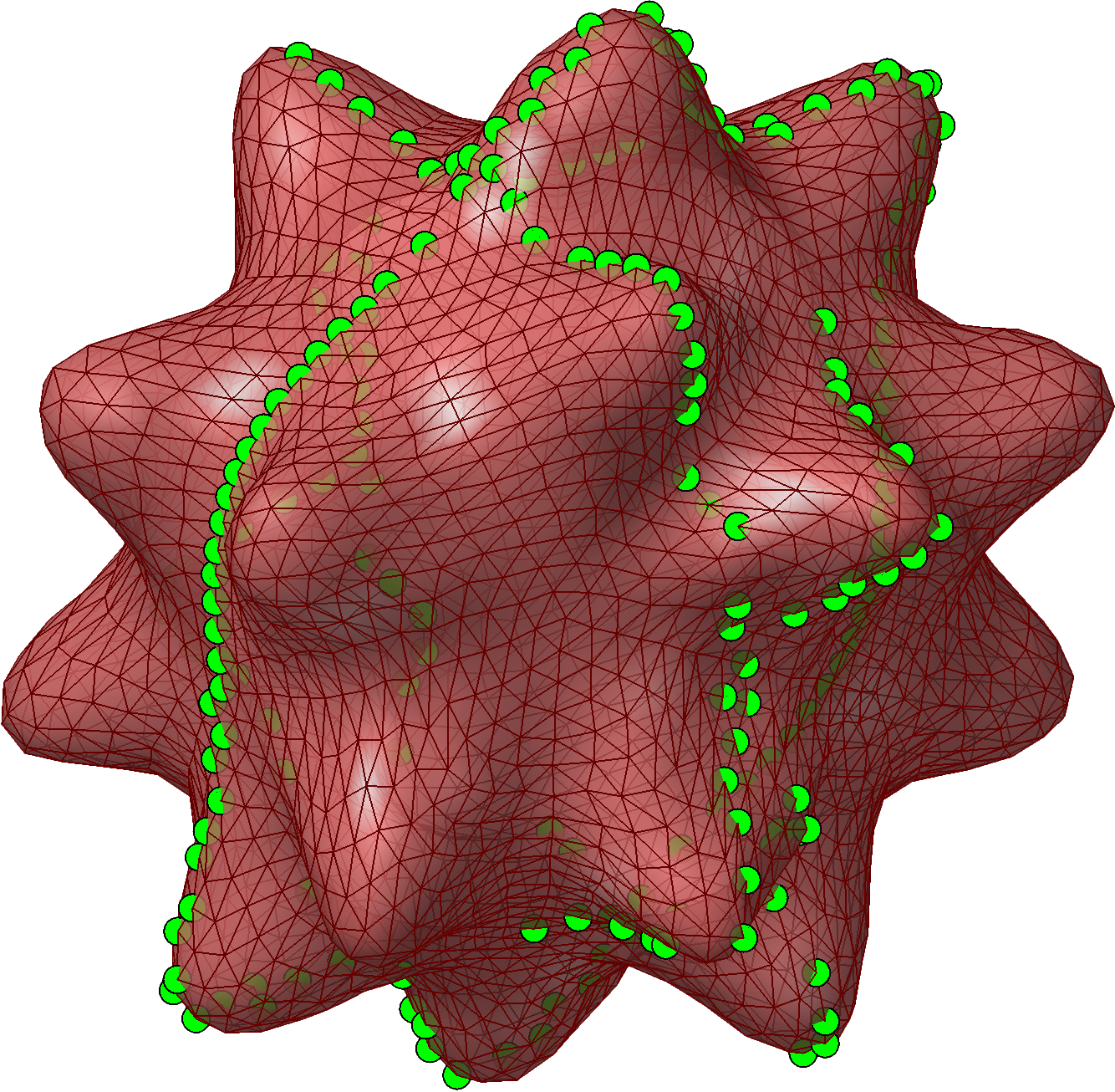}} & \adjustbox{valign=m}{\includegraphics[width=0.1\textwidth]{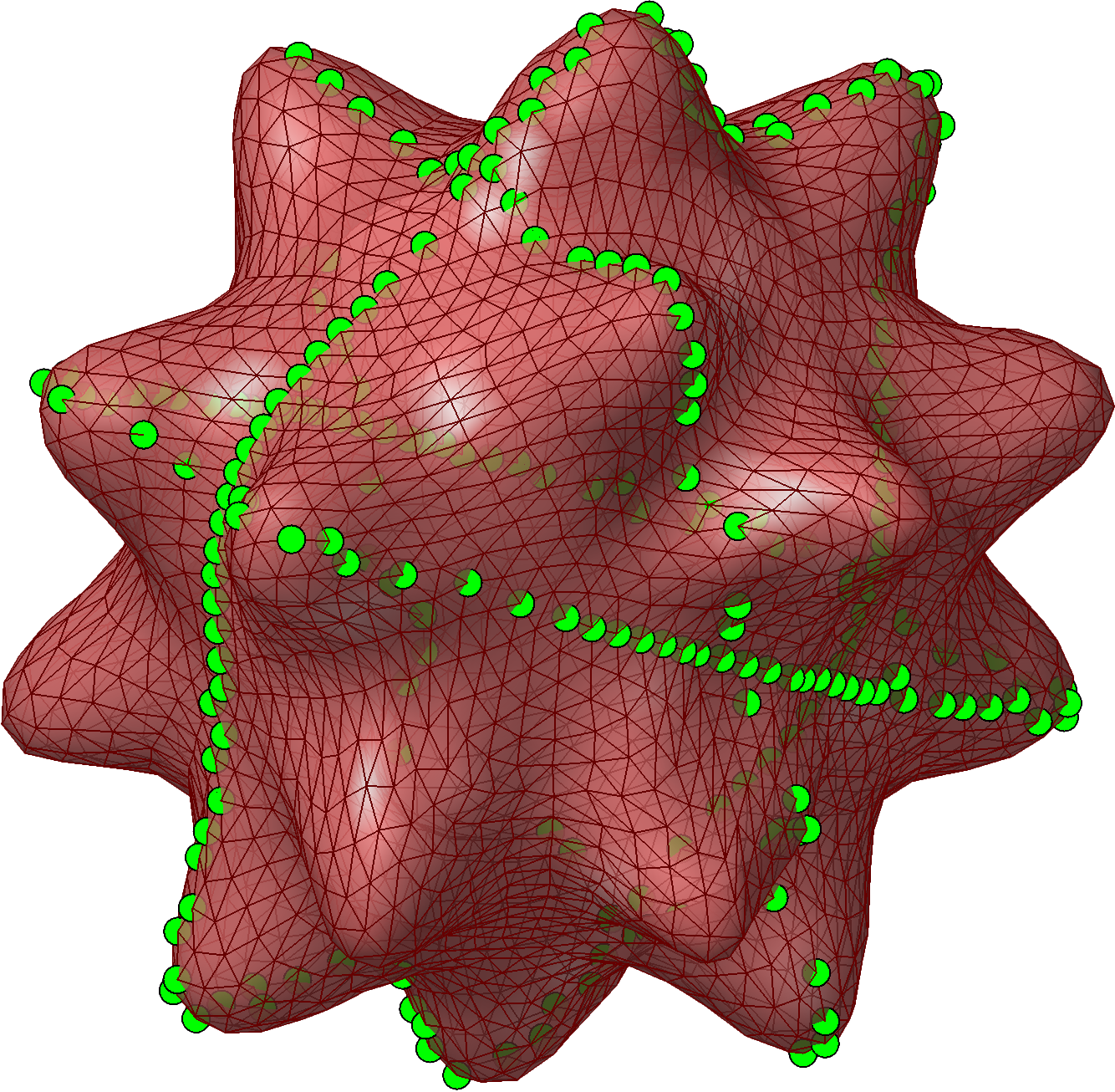}} & \adjustbox{valign=m}{\includegraphics[width=0.1\textwidth]{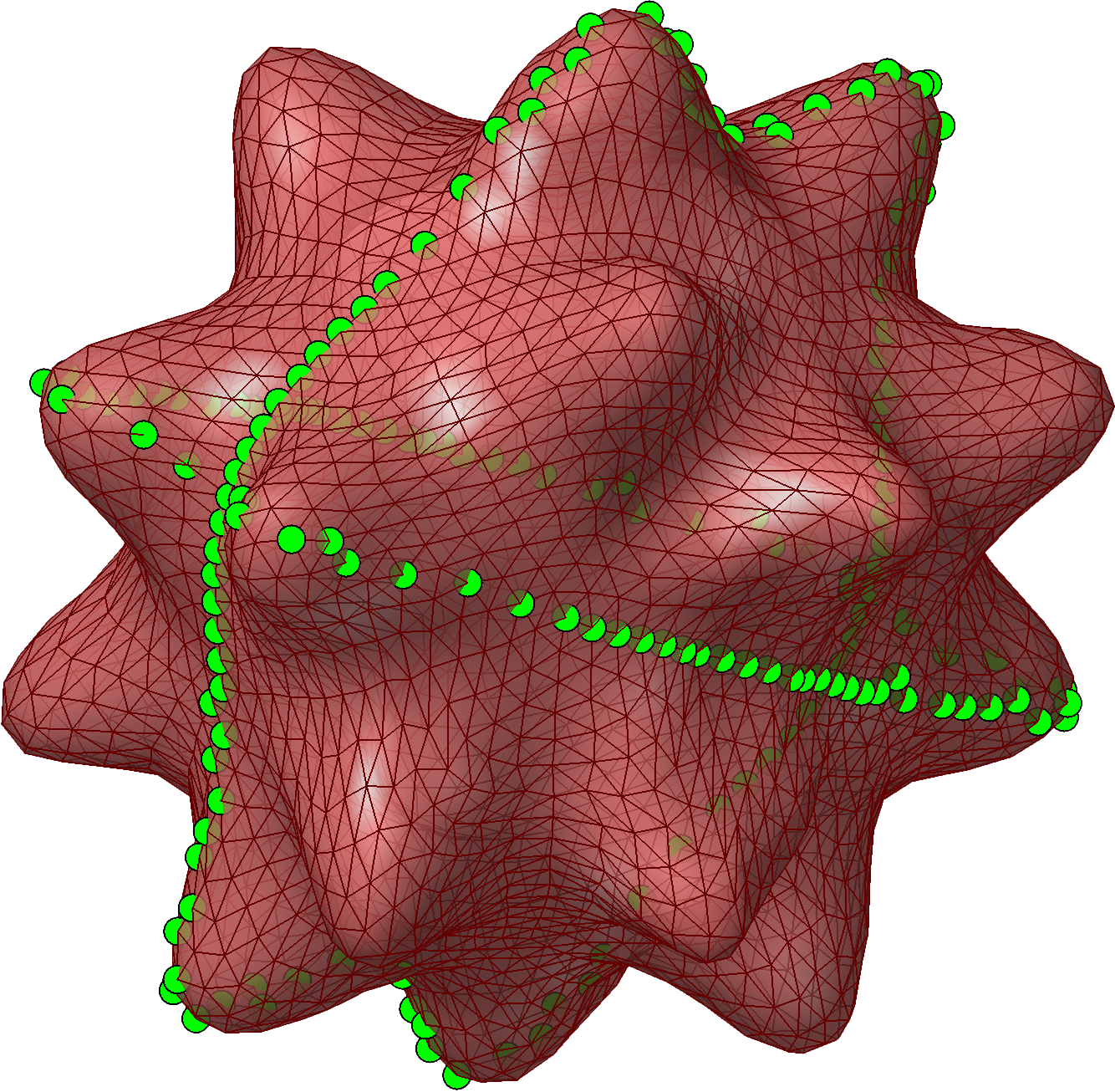}} & \adjustbox{valign=m}{\includegraphics[width=0.1\textwidth]{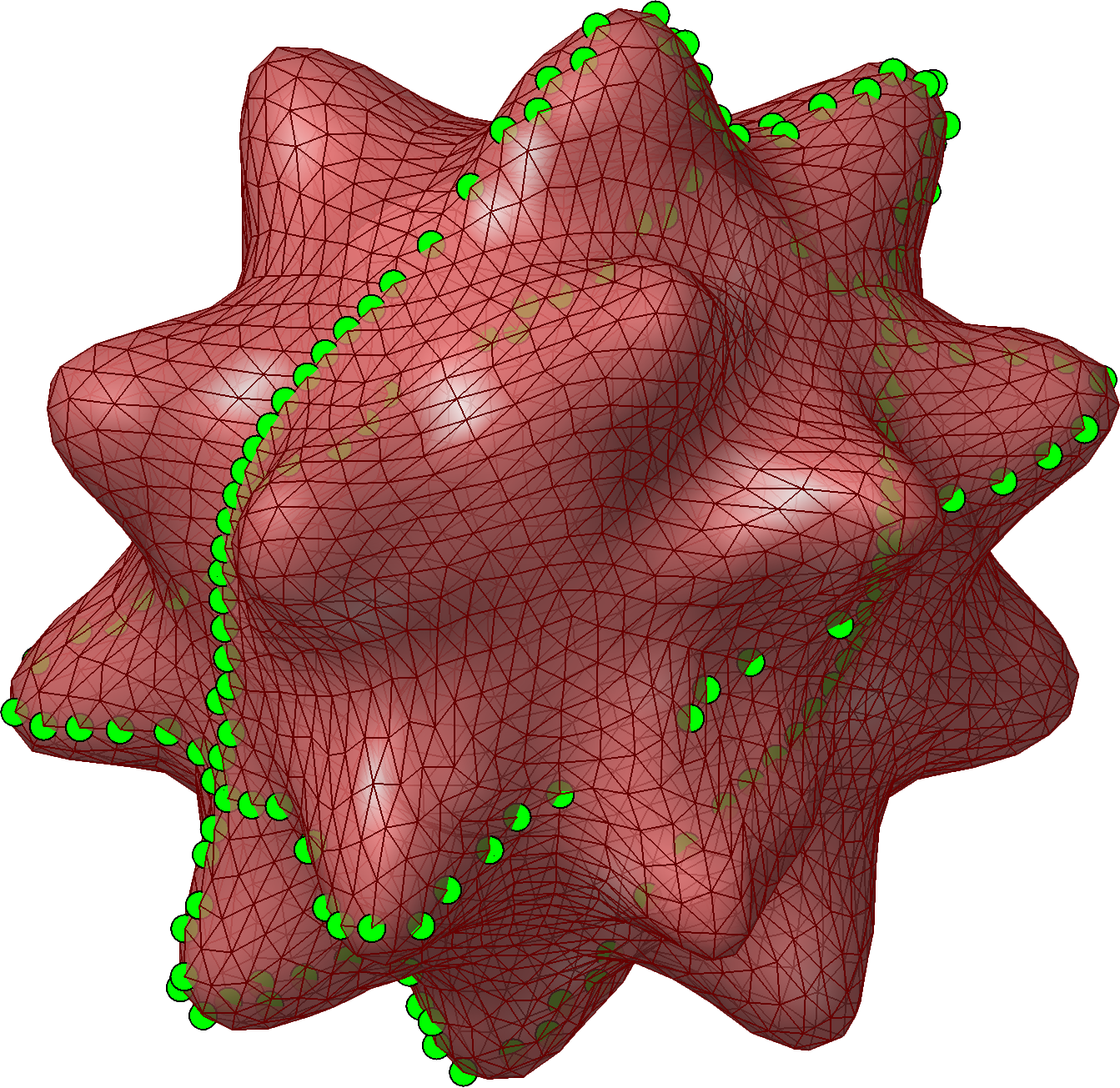}} \\ [20pt]
Sample number & 340 & 205 & 182 & 227 & 173 & 177 \\
\bottomrule
\end{tabularx}
\end{table}

\biboptions{sort&compress}

\end{document}